\documentclass[11pt]{article}

\usepackage[final]{acl}

\usepackage{times}
\usepackage{latexsym}
\usepackage{amssymb}
\usepackage{booktabs} 
\usepackage{tabularx}
\usepackage{listings}
\usepackage{colortbl}
\usepackage{wrapfig}
\lstdefinestyle{prompt}{basicstyle=\footnotesize,breaklines=true,%
  breakindent=0pt,breakatwhitespace=true,columns=fullflexible,%
  keepspaces=true,aboveskip=1pt,belowskip=1pt,frame=none,xleftmargin=0pt}

\usepackage[T1]{fontenc}

\definecolor{salmon}{RGB}{247, 173, 150}

\newlength{\barunit}
\usepackage[utf8]{inputenc}

\usepackage{microtype}

\definecolor{accentpurple}{RGB}{155,120,220}
\colorlet{calloutbar}{accentpurple!85!black}
\colorlet{calloutbg}{accentpurple!12}
\usepackage[most]{tcolorbox}
\newtcolorbox{takeaway}[1]{%
  enhanced, breakable,
  colback=calloutbg, frame hidden, boxrule=0pt, arc=2pt,
  borderline west={3pt}{0pt}{calloutbar},
  left=12pt, right=12pt, top=8pt, bottom=8pt,
  fontupper={\linespread{1}\small\selectfont},
  before upper={\textbf{\color{black}#1}\par\smallskip}%
}

\usepackage{tcolorbox}

\definecolor{openaigreen}{RGB}{58,143,106}     
\definecolor{googleblue}{RGB}{74,127,212}      
\definecolor{anthropicorange}{RGB}{192,83,43}  
\definecolor{othersgray}{RGB}{85,85,85}
\newlength{\modelbarlen}
\newcommand{\modelbar}[3]{\rlap{{\color{#2}\rule[-0.55ex]{#1\modelbarlen}{2.1ex}}}\hspace{2pt}#3}

\usepackage{inconsolata}
\usepackage{enumitem}
\usepackage{graphicx}

\title{Critical Model Lifecycle Dependency for Scientific Progress}

\author{
    \textbf{Wenchao Dong}$^{1}$~~~~
    \textbf{Marcelo Sartori Locatelli}$^{2}$~~~~
    \textbf{Virgilio Almeida}$^{3}$~~~~
    \textbf{Meeyoung Cha}$^{1,4}$\\
    $^{1}$MPI-SP, Germany~~~~
    $^{2}$Cornell University, USA~~~~
    $^{3}$UFMG, Brazil~~~~
    $^{4}$KAIST, South Korea\\
    \small{\textbf{Correspondence:} \href{mailto:mia.cha@mpi-sp.org}{mia.cha@mpi-sp.org}}
}

\begin{document}
\maketitle


\begin{abstract}
Large language models (LLMs) have become central instruments throughout the NLP research cycle, yet the most capable are built by a few companies and can be modified or withdrawn at any time.
In this paper, we audit this dependence empirically across 74,029 published ACL Anthology papers (2018--2026), extracting every model each paper uses, together with the authors' affiliations and countries.
Our results show that, since ChatGPT's release, nearly half of all papers run a closed model from OpenAI, Google, or Anthropic, with OpenAI models appearing in 95\% of them.
These instruments are already disappearing.
Among closed model papers, 71.6\% use at least one model version that has since been deprecated, and a third use one that has already been shut down, so their experiments can no longer be re-executed through any official interface.
With no standard protocol for reporting model usage, only 14\% of closed model mentions include an exact snapshot version, making the decay even harder to audit.
Model usage also diffuses unevenly, stratified by institution, venue, geography, and wealth.
We argue that reproducibility must become a reported property of every paper.
To this end, we propose a model-usage checklist covering each model's exact version, access platform, lifecycle status, replicability-after-retirement plan, and cost, serving as a codebook for authors and reviewers of LLM-mediated science.
\end{abstract}


\section{Introduction}

Large generative models have become ubiquitous in scientific research, nowhere more so than in natural language processing (NLP).
Since the introduction of BERT in 2018 and the release of ChatGPT in 2022~\cite{zhao2026surveyllms}, language models have reshaped both what the field studies and how it studies it.
Beyond serving as objects of study, large language models (LLMs) now accelerate research itself.
Agentic systems automate literature synthesis~\cite{asai2026synthesizing}, code generation~\cite{Simone2026Who}, scientific discovery~\cite{ghareeb2026scientific}, and even end-to-end AI research~\cite{lu2026towards}, while informal LLM assistance touches every step of scientific work, from designing experiments to drafting manuscripts and writing reviews.
The gains are measurable: authors using LLM assistance publish over 30\% more~\cite{kusumegi2025scientific}, AI-related submission volumes keep climbing~\cite{sha2026ai}, and LLM-related arXiv categories received 2--16$\times$ more submissions after 2023~\cite{movva2024topics}.

This acceleration, however, hides a structural risk~\cite{Fleisher2026Inductive}, as the empirical work of NLP now runs on instruments that the field does not control.
Most are commercial products served from a model provider's infrastructure. Hence, they may be silently updated~\cite{chen2024chatgpt}, deprecated, or shut down entirely~\cite{ma2024schrodinger}.
Such retirements are routinely carried out as the cost of keeping models available grows with the number of models served~\cite{anthropic2025deprecation}.
An update changes the instrument beneath published results, a shutdown makes re-execution impossible, and routinely omitted version identifiers or access dates leave it unclear which model a result rests on.
Because reproducibility\footnote{Throughout this work, \emph{reproducibility} refers to the re-execution of original experiments. We do not examine the falsifiability of the claims derived from them.} is a core principle of science~\cite{Collins1985-COLCOR}, claims that cannot be re-examined become an unreliable foundation to build on~\cite{ioannidis2005most}.

Even when the instruments remain available, access to them is unequal.
Research is centralizing around a few key organizations~\cite{fan2024bibliometric}, while the costs of compute~\cite{lehdonvirta2024compute} and proprietary APIs~\cite{ahia2023all} push institutions with fewer resources, particularly in the Global South, toward the margins.
Furthermore, review expectations calibrated to the newest models can bury otherwise sound work.

In this paper, we audit how extensively the field builds on a limited set of language models and how accessible those models remain over time and across institutions and countries.
From a fixed snapshot of the ACL Anthology, we extract every model used by each of the 74,029 papers published between 2018 and 2026, together with the authors' affiliations and countries, and link the closed models to their providers' official lifecycle records.
Three findings emerge.

\begin{enumerate}[itemsep=1pt,topsep=2pt]
    \item \textbf{NLP research runs on a handful of models.} Nearly half of the papers published since 2023 use a closed model from OpenAI, Google, or Anthropic. OpenAI models appear in 95\% of these papers, and all other commercial providers combined reach less than a tenth of OpenAI's reach.
    
    \item \textbf{Scientific evidence is already expiring.} As of our June 2026 snapshot, 71.6\% of closed model papers use at least one model version that has since been deprecated, and a third use models that have already been shut down, so those experiments can no longer be re-executed through any official interface. Only 14\% of closed model mentions include an exact snapshot version, making the decay even harder to audit.

    \item \textbf{Access to frontier models is unequal.} LLM types appear in a highly concentrated manner and diffuse only gradually across institutions and countries. Usage further splits by venue, geography, and wealth, with richer countries migrating fastest from classic pretrained models to open-weight LLMs.

\end{enumerate}

We argue that large language models now function as critical infrastructure for the field~\cite{rinaldi2001identifying}.
When a field's empirical claims rest on a small number of externally controlled models, the lifecycle, pricing, and access policies of those models determine what science can be done, verified, and built upon.
To keep the benefits of frontier models while ensuring scientific results are verifiable, we propose a \textbf{model-usage checklist} that records each model's exact version, access platform, lifecycle status, replicability-after-retirement plan, and cost.
We offer concrete research recommendations and discuss how the research community should address this dependence in advancing reproducible science.
Our experimental data are publicly shared on GitHub.\footnote{\url{https://github.com/model-depend/lifecycle}}


\section{Background}

\paragraph{Reproducibility and replicability in science.}

Reproducibility is fundamental to scientific research. 
Merton's norm of universalism~\cite{merton1942sociology} states that scientific claims should be evaluated by impersonal criteria, implying that any scientist should, in principle, be able to verify existing results given adequate resources~\cite{Collins1985-COLCOR}. 
Post-academic science, shaped by commercial interests and shifting norms~\citep{ziman2001real}, often fails to meet this ideal. 
When science faces reproducibility crises and a substantial fraction of published findings become non-replicable~\citep{ioannidis2005most,open2015estimating}, this failure (often driven by hype and economic incentives) undermines public trust in scientific research.
Computational science should be inherently easier to reproduce since data, code, and executables suffice without dedicated equipment~\cite{peng2011reproducible}, yet intellectual property concerns and documentation effort remain barriers~\cite{stodden2010scientific}.

\paragraph{Reporting standards and checklists in AI/NLP venues.}

Concerns over scientific practice in AI have produced a series of disclosure standards, including model cards~\cite{mitchell2019model} and analogous documentation for datasets and supplementary resources~\citep{bender2018data,gebru2021datasheets}. 
Worries about unrepeatable experiments~\cite{dodge2019show} and irresponsible data usage~\cite{rogers2021just} led the community to enforce author-completed checklists at submission time. 
Following the NeurIPS reproducibility program~\cite{pineau2021improving}, other venues adopted similar instruments, including the ARR Responsible NLP Research Checklist~\cite{arr2024responsible} and the AAAI Reproducibility Checklist~\cite{aaai2026reproducibility}. 
Despite the effort, rapid progress in technology raises new challenges these measures do not yet cover. 
To fill the gap, initiatives like the GUIDE-LLMs checklist~\cite{feuerriegel2026reporting} offer more concrete guidance on transparency, reproducibility, and the ethical use of LLMs in behavioral science. 
We further contribute to this effort by documenting model lifecycles in scientific reproducibility.

\paragraph{Reproducibility for proprietary and open-weight models.} 
Sharing data, code, and executables no longer guaranties reproducibility. 
Responses of the same proprietary model may drift substantially over time~\cite{chen2024chatgpt}, and models can become unavailable at any moment~\cite{ma2024schrodinger, trisovic2026the}, harming reproducibility even when all artifacts are shared. 
\citet{angermeir2025reflections} found that none of 85 LLM-centric studies relying on OpenAI models could be fully reproduced, with deprecation and version dependencies among the contributing factors. 
Opaque, setup-sensitive evaluation practices that often lack statistical rigor compound the problem~\citep{biderman2024lessons,la2024language}. 
Open-weight LLMs mitigate some concerns, since works built on them are, in principle, reproducible, but the models themselves are not, as training data, code, and key details remain hidden, leaving them usable yet not scrutinizable~\cite{widder2024open}.

\paragraph{Compute and costs.} 

The current compute landscape further limits reproducibility and open AI research~\cite{widder2024open}.
Efficiency is neglected in favor of state-of-the-art results, raising environmental and monetary costs and making NLP research prohibitively expensive~\cite{schwartz2020green}, a problem intensified by unequal access to compute~\cite{lehdonvirta2024compute}. 
Despite calls for greener and more democratic AI~\citep{strubell2019energy,ahmed2020democratization}, trends still point to a concentration of power and growing compute requirements for participating in AI research~\citep{ahmed2023growing,besiroglu2024compute}. 
This deepens critical infrastructure interdependency, as the field becomes entangled with a few model providers whose decisions propagate directly into science, shaping what gets researched~\cite{rinaldi2001identifying}.



\section{Methodology}
\label{sec:method}

\subsection{Corpus and Feature Extraction}

\paragraph{Corpus.}
We analyze a fixed snapshot (2026-06-24) of the ACL Anthology, restricting ourselves to papers published in 2018 or later, a period that extends from pretrained language models (e.g., BERT) to current LLMs~\cite{zhao2026surveyllms}.
We limit our audit to published works rather than preprints as a quality control on which conclusions enter the scientific record.
We downloaded the official camera-ready PDFs of the entire \textbf{74,029} papers and extracted the text. 
Based on the release of ChatGPT, we split this span into the pre-LLM era (2018--2022; 32,010 papers) and the LLM era (2023--2026; 42,019 papers).
See Appendix~\ref{sec:appendix_dataset_details} for the preprocessing details.

\paragraph{Feature extractions.}
We extract all the models used in each paper by prompting Gemma4-31B-it with the entire content after stripping bibliography sections, yielding 689,588 model mentions.
Extracted model mentions are normalized across spelling variants by lowering cases and unifying dashes and spaces (e.g., we treat GPT 4o and gpt-4o as the same name).
To obtain the distinct models used for each paper, a generic name was considered instead of a specific name (e.g., gpt-4o rather than gpt-4o-2024-08-06).

We used rule-based mapping to classify each model into one of the following three types: (1) classic pretrained language models (PLMs) of the BERT era; (2) closed LLMs from commercial providers; and (3) open-weight LLMs. 
For open-weight models, we extract model sizes and standardize parameter counts to billions, parsed from the model name when available (e.g., 410M $\rightarrow$ 0.41B; 8$\times$7B $\rightarrow$ 56B), as in Appendix~\ref{sec:appendix_model_extraction}.

We validated the model extraction in two ways.
First, we manually checked every extracted model against the PDFs for a stratified sample of 100 papers across era, venue, and model type; across 425 extracted models, the average accuracy is 94.6\%.
Second, we repeated the extraction over all papers with a second model, Qwen3.8-27B, using the same prompt, resulting in an average agreement of 96.7\% between the two models' extractions (Cohen's $\kappa=0.93$).

We extract the author list with associated affiliations and country information for each author using Gemma4-31B-it.
Institutions are standardized according to the Research Organization Registry~\cite{ror} to their full names in order to merge across publications, as described in Appendix~\ref{sec:appendix_country_extraction}.

\paragraph{Pairing with model lifecycles.}

For the three major commercial model providers, we compile the release dates, deprecation announcements, and shutdown dates of every used model version from the providers' official websites.
We assign each model usage a lifecycle status as of the corpus snapshot date: ongoing, deprecated, or shut down.
Because mentions that are only product names cannot be resolved, all the following analyzes remain as lower bounds of real effects. Details are in Appendix~\ref{sec:appendix_lifecycle}.

\subsection{Measurements}

\paragraph{Model-type usage per paper.}

Based on extracted models for each paper, we derive binary usage flags for each paper and model type: whether the paper used at least one closed LLM from the three major providers, at least one open-weight LLM, and at least one classic PLM.
These measure model adoption, not intensity, as we do not measure how many models in each type are used for robustness reasons (see discussions in Appendix~\ref{sec:appendix_model_extraction}).
We further classify each open-weight model into one of three size categories ($\leq$10B, 10--70B, $>$70B).
A paper can fall into more than one category, as it may use models of different sizes.

We further calculate model usage per venue, institution, and country.
Venue counts equal paper counts, as every paper belongs to exactly one venue.
Institutions and countries were measured through authors' affiliation lists.
Each paper splits one unit of credit equally among its authors, and each author's share is equally distributed among their affiliations and those affiliations' countries.
The split credits sum to at most one per paper since some affiliations are unresolvable, as described in Appendix~\ref{sec:appendix_gini}.

\paragraph{Usage concentration: Gini index.}

To quantify how unequally the model types used are distributed across publishing institutions or countries in a year, we calculate the Gini index.
The index is zero when every published source carries an equal share of the type's usage and approaches one as usage concentrates in a single source.

\paragraph{Economic imbalance.}

We consider the economic imbalance of model usage by pairing each country or region's per-year model usage with its GDP per capita through the World Bank's World Development Indicators~\cite{worldbank}.
We fit per-year regression models of log-scale GDP per capita on publication counts and on each model type's usage share. See Appendix~\ref{sec:appendix_gdp} for details.


\section{Results}

\subsection{Model Concentration}
\label{sec:model_concentration}

\begin{table}[t]
\centering
\small
\renewcommand{\arraystretch}{1.15}
\setlength{\modelbarlen}{2.9cm}%
\resizebox{\columnwidth}{!}{%
\begin{tabular}{lr@{\hspace{7pt}}!{\color{gray!100}\vrule}@{\hspace{7pt}}lr}
\toprule
\multicolumn{2}{@{}l}{\textit{OpenAI (19{,}619)}} & \multicolumn{2}{@{}l}{\textit{Anthropic (3{,}383)}} \\
\modelbar{1.000}{openaigreen!62}{GPT-4} & 6{,}513 & \modelbar{0.149}{anthropicorange!62}{Claude-3.5-Sonnet} & 972 \\
\modelbar{0.998}{openaigreen!52}{GPT-4o} & 6{,}502 & \modelbar{0.084}{anthropicorange!52}{Claude} & 548 \\
\modelbar{0.557}{openaigreen!42}{ChatGPT} & 3{,}625 & \modelbar{0.052}{anthropicorange!42}{Claude-3.7-Sonnet} & 337 \\
\modelbar{0.539}{openaigreen!32}{GPT-3.5-turbo} & 3{,}513 & \modelbar{0.046}{anthropicorange!32}{Claude-Sonnet-4} & 302 \\
\modelbar{0.459}{openaigreen!22}{GPT-3.5} & 2{,}988 & \modelbar{0.043}{anthropicorange!22}{Claude-3-Opus} & 278 \\
\addlinespace[2pt]
\midrule
\addlinespace[2pt]
\multicolumn{2}{@{}l}{\textit{Google (4{,}793)}} & \multicolumn{2}{@{}l}{\textit{Others (1{,}716)}} \\
\modelbar{0.144}{googleblue!60}{Gemini} & 935 & \modelbar{0.087}{othersgray!68}{Alibaba} & 566 \\
\modelbar{0.114}{googleblue!50}{Gemini-2.5-Pro} & 740 & \modelbar{0.061}{othersgray!58}{Zhipu AI} & 397 \\
\modelbar{0.103}{googleblue!40}{Gemini-2.5-Flash} & 670 & \modelbar{0.038}{othersgray!48}{xAI} & 247 \\
\modelbar{0.098}{googleblue!30}{Gemini-1.5-Pro} & 640 & \modelbar{0.031}{othersgray!38}{ByteDance} & 203 \\
\modelbar{0.078}{googleblue!20}{Gemini-2.0-Flash} & 510 & \modelbar{0.026}{othersgray!28}{Moonshot AI} & 169 \\
\bottomrule
\end{tabular}%
}
\caption{Most referenced closed models.}
\label{tab:closed_model_usage}
\end{table}

\paragraph{Closed model concentrates on a single provider.}
A total of 20,575 papers (i.e., 27.8\% of the corpus; 47.8\% in the LLM era) use at least one closed model from the three major providers.
OpenAI models appear in 19,619 (95\%) of these papers (Table~\ref{tab:closed_model_usage}), nearly four times the reach of Google and six times that of Anthropic.
All remaining commercial providers combined reach 1,716 papers, with the largest, Alibaba, roughly 35 times less often than OpenAI.

\begin{wrapfigure}{r}{0.28\textwidth}
\centering
\includegraphics[width=\linewidth]{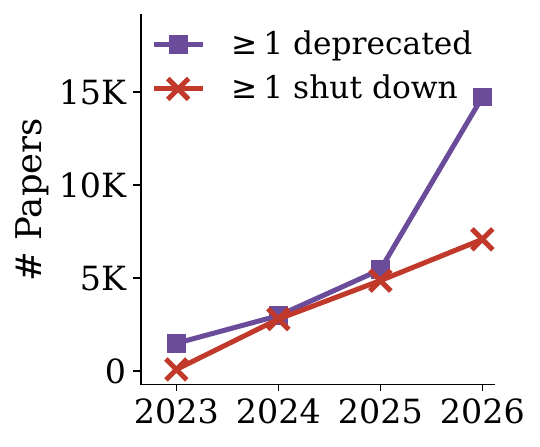}
\caption{Cumulative status.}
\label{fig:status_cumulative}
\end{wrapfigure}

\paragraph{Papers depend on deprecated models.}

Of the 20,575 papers using a major-provider closed model, 17,829 (86.7\%) match versions with official lifecycle dates.
Across all 56,359 closed-model mentions, 40,404 (71.7\%) resolve to a dated release, yet only 7,909 (14.0\%) specify an exact snapshot.
The remainder rely on family-level names (such as ``GPT-4'') that inherit official retirement dates because providers typically discontinue entire model lines at once.
Cumulative counts of papers affected by deprecations and shutdowns are shown in Figure~\ref{fig:status_cumulative}.
As of June 2026 snapshot, 14,723 papers (71.6\% of those using major-provider closed models) rely on at least one deprecated model, and 7,067 (34.3\%) include at least one shut-down model.
For 1,348 papers, \emph{every} major-provider closed model used has been shut down, leaving their closed-model experiments impossible to re-run through any official interface.
This trend is accelerating. The number of papers citing deprecated models nearly tripled between late 2025 and mid-2026.

\paragraph{Open-weight LLMs concentrate on small sizes.}

\begin{figure}[t]
  \centering
  \includegraphics[width=\linewidth]{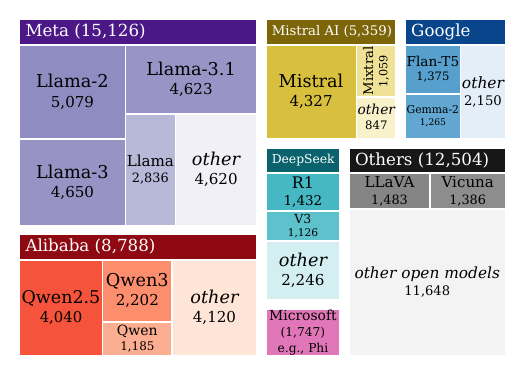}
  \caption{Open-weight LLM usage, with block area proportional to the number of papers using each model. Group headers report unique papers; model counts overlap, as a single paper can use multiple models.}
  \label{fig:open_model_treemap_small}
\end{figure}

Across the corpus, 22,883 papers (30.9\% overall and 53.6\% of LLM-era studies) use at least one open-weight LLM, evaluating a mean of 4.6 distinct models per paper (median 3).
In 20,191 of these papers (88.2\%), at least one open-weight model maps to a known size; across the distinct sized models in each paper, the mean size is 24.9 billion parameters ($SD=58.9$).
The 7--8B class accounts for 44\% of sized open-weight mentions, with a median model size of 8B and a large mean-median gap suggesting a strongly right-skewed distribution.
Table~\ref{tab:open_size_ranked} lists the ten most-used model sizes: the 7B and 8B classes lead by a wide margin (12,198 and 8,826 papers), followed by 70B as the most popular large size.
Figure~\ref{fig:open_model_treemap_small} breaks usage down by developer: Meta dominates with 15,126 papers, driven by the Llama family, followed by Alibaba's Qwen models (8,788), Mistral AI (5,359), Google (4,372), and DeepSeek (3,815).

\begin{table}[t]
\centering
\resizebox{\columnwidth}{!}{%
\begin{tabular}{@{}lrl@{\hspace{1.6em}}|lrl@{}}
\toprule
\textbf{Size} & \textbf{\#Paper} & \textbf{Example Models} & \textbf{Size} & \textbf{\#Paper} & \textbf{Example Models} \\
\midrule
7B & 12{,}198 & Mistral, Llama-2 & 32B & 2{,}109 & Qwen2.5, Qwen3 \\
8B & 8{,}826 & Llama-3.1, Llama-3 & 72B & 1{,}968 & Qwen2.5, Qwen2.5-VL \\
70B & 4{,}605 & Llama-3, Llama-3.1 & 14B & 1{,}875 & Qwen2.5, Qwen3 \\
13B & 3{,}694 & Llama-2, Vicuna & 9B & 1{,}467 & Gemma-2, GLM-4 \\
3B & 2{,}728 & Llama-3.2, Qwen2.5 & 1B & 1{,}274 & Llama-3.2, Pythia \\
\bottomrule
\end{tabular}%
}
\caption{The ten most-used open-weight model sizes, ranked by the number of papers using each size at least once. Examples show the most-used model families.}
\label{tab:open_size_ranked}
\end{table}

\subsection{Model Imbalance}
\label{sec:results_imbalance}

\begin{figure}[t]
  \centering
  \includegraphics[width=.88\linewidth]{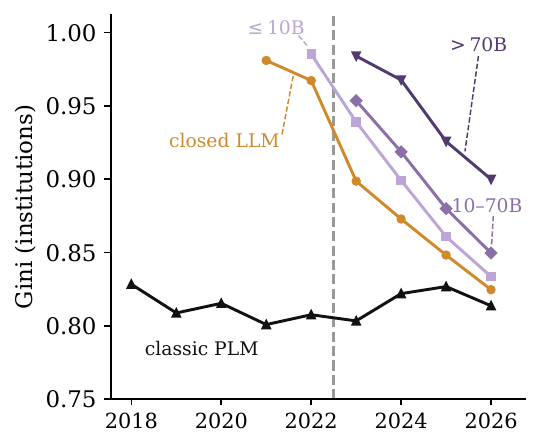}
  \caption{Concentration of model-type usage across publishing institutions, measured by the per-year Gini index. Open-weight LLMs are split by parameter size. The dashed line separates the pre- and post-LLM eras.}
  \label{fig:ineq_gini_by_year}
\end{figure}

\paragraph{LLM usage starts concentrated, then diffuses.}
Figure~\ref{fig:ineq_gini_by_year} tracks how unequally each model type's usage is distributed over publishing institutions over years.
Classic PLM usage has been stable near a Gini of 0.81 since 2018.
Every LLM type begins at near-total concentration, with closed LLMs at 0.98 in 2021 and each open-weight tier at 0.95--0.99, before diffusing across institutions over time. By 2026, closed LLMs and small ($\leq$10B) open models reach 0.82--0.83.
No LLM series has come down to the classic PLM level yet, and concentration stays ordered by scale: the larger the open-weight tier, the more unequal its use, with $>$70B models still at 0.90 in 2026.
A similar trend can be observed in Figure~\ref{fig:ineq_gini_by_year_country} for Gini index by country.

\begin{figure}[t]
  \centering
  \includegraphics[width=.95\linewidth]{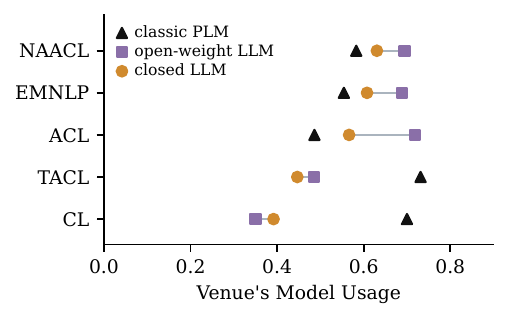}
  \caption{Share of each venue's papers model usage.}
  \label{fig:ineq_venue_flagship}
\end{figure}

\begin{figure}[t]
  \centering
  \includegraphics[width=\linewidth]{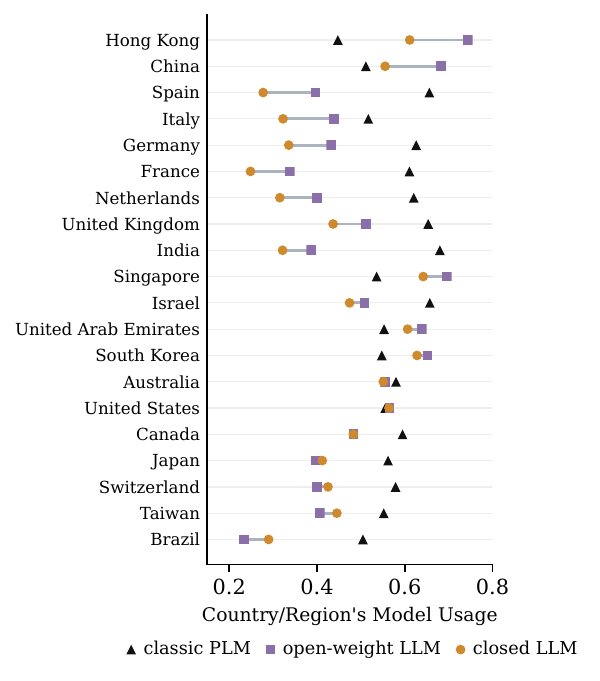}
  \caption{Share of each country/region's model usage.}
  \label{fig:ineq_country_rates}
\end{figure}

\paragraph{Adoption by venue and geography.}

In the LLM era, across five selected venues (Figure~\ref{fig:ineq_venue_flagship}; extended venues in Appendix Figure~\ref{fig:app_ineq_venue}), the three main conferences share a similar trend, in which open-weight LLMs lead (69--72\% of papers), followed by closed LLMs (57--63\%), then classic PLMs. 
The two journals invert this order: TACL and CL remain PLM-first (73\% and 70\%) with far lower LLM usage (35--49\%), possibly due to their longer submission-to-publication timeline (see Table~\ref{tab:app_venue_delay}).

Among the 20 countries and regions carrying the most paper credit (Figure~\ref{fig:ineq_country_rates}), classic PLMs remain the single most-used type in 14, including all seven European countries shown, whereas China, Hong Kong, Singapore, South Korea, the United States, and the United Arab Emirates are LLM-first. 
Hong Kong and China especially lean toward open weights (74\% and 68\% of papers, 12--13\,pp above their closed shares), while the United States uses all three types at essentially the same rate ($\approx$56\%). 
Only four countries use closed models more than open ones; France has the lowest closed LLM share of the panel (25\%) and Brazil the lowest open-weight share (23\%).
See Table~\ref{tab:country_rates_full} for extended country results.

\paragraph{Model choice tracks economic wealth.}

\begin{figure}[t]
  \centering
  \includegraphics[width=.8\linewidth]{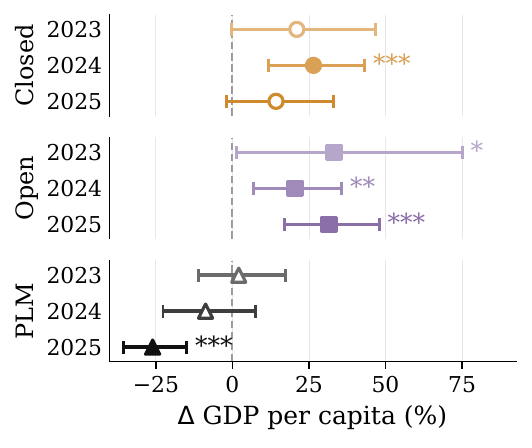}
  \caption{Coefficients show the estimated change in GDP per capita per 10-pp increase in a country's usage share of each model type, fit separately per publication year; error bars are 95\% confidence intervals, and solid markers indicate statistically significant. $^{*}\,p<0.05$, $^{**}\,p<0.01$, $^{***}\,p<0.001$.}
  \label{fig:gdp_coefficient}
\end{figure}

When regressing GDP per capita on each country's model usage (Figure~\ref{fig:gdp_coefficient}), a 10-pp-higher open-weight-LLM share is associated with 20--33\% higher GDP per capita, significant in every year and strongest in 2025 (+31.6\%, $p<0.001$). 
Classic PLM shares carry no signal in 2023--2024, but by 2025 a 10-pp-higher PLM share marks countries 26\% poorer ($p<0.001$), suggesting PLM reliance concentrated in lower-income countries and richer countries migrated to open-weight LLMs faster. 
Closed LLM shares are the weakest correlate of wealth, consistent with paid APIs requiring no local compute investment. 
These associations are correlational rather than causal; full estimates are in Appendix~\ref{sec:appendix_gdp}.



\section{Discussions}

\subsection{Reproducibility for Scientific Knowledge}
\label{sec:disc_repro}

At the current pace of model deprecation, failing to address it renders the conclusions derived from these models irreproducible by default.
Using frontier models is a legitimate, often necessary, scientific practice: for papers that probe newly released systems, the newest model is not a design choice but the object of study, and the fact that a third of closed model papers already depend on a shut-down model (Figure~\ref{fig:status_cumulative}) partly reflects how quickly the community adopts new technology.
The problem arises downstream, when empirical claims outlive the instruments that produced them.
Once a provider retires an endpoint, published experiments cannot be re-executed through any official interface~\cite{angermeir2025reflections}; even while a model is served, silent updates and inference nondeterminism erode the evidential value of results built on it~\cite{chen2024chatgpt, song2025greedy}.

\subsection{Difficulty of Auditing Model Dependency}
\label{sec:disc_audit}

Measuring model dependence is itself challenging. 
Our estimates are lower bounds by construction: only 14\% of major-provider mentions pin an exact model snapshot, and nearly a third cannot be resolved to any dated version even at the family level, often because they use only a product name (e.g., ``ChatGPT'').
The access route is also not clearly documented.
The same model may be reached through the provider's API, a cloud endpoint, or a third-party aggregator, and these routes can differ in version, defaults, and even in which model is actually served~\cite{cai2025getting}. 
No major venue checklist asks authors to record this route (Table~\ref{tab:checklist_mapping}). 
Black-box access already limits what an audit can establish~\cite{casper2024blackbox,la2024language}, and black-box access through an unnamed intermediary makes even the identity of the instrument unverifiable. 
Recording this access information at publication time would make future audits of model reliance more tractable.

\subsection{Request Blocking for Specific Domains}

Providers can decide which fields may use their models.
Deployment safeguards now refuse or filter requests in specific scientific domains.
Anthropic's AI Safety Level~3 protections restrict workflows judged related to chemical, biological, radiological, or nuclear risk~\cite{anthropic2025asl3,anthropic2025nuclear}, and OpenAI documents comparable safeguards~\cite{openai2026gpt55systemcard}.
These safeguards are designed to mitigate real misuse risks, but they are typically enforced by smaller classifiers at inference time, leading to potential over- or under-blocking.
Beyond misclassification, such filtering also restricts researchers in these domains from leveraging cutting-edge AI systems for their own scientific progress.
The model concentration documented in our results (\S\ref{sec:model_concentration}) further indicates that no adequate substitute exists.
Although structured access and legal safe harbors for vetted researchers have been advocated~\cite{longpre2024safe}, access to unfiltered models remains limited to large corporations.
Domain-specific safety filters are thus a further axis of dependence.

\subsection{Model Concentration and Imbalance}

With OpenAI models in 95\% of closed model papers, much of what the field knows about ``LLMs'' is in practice knowledge about one provider's product, such as tokenization and its cost structure~\cite{ahia2023all}, cultural skew~\cite{tao2024cultural}, and safety filtering, which propagate into thousands of studies as if they were general properties, and errors correlate across the literature in the manner of algorithmic monoculture~\cite{bommasani2022picking}. 
Concentration also creates a single point of failure, as one pricing or access decision simultaneously alters the experimental conditions of a large share of the field's studies. 
The diffusion we observe (Figure~\ref{fig:ineq_gini_by_year}) shows falling entry costs, yet access remains stratified by model size. 
The largest open-weight tier is still more concentrated than every other model type in 2026, consistent with the compute divide~\cite{besiroglu2024compute}. 
Model lifecycle and accessibility thus act together: which models remain available, and which a community can afford, determine what evidence it can produce and reproduce.

\subsection{Recursive Pipeline}

Model dependence extends beyond individual experiments into the entire process of scientific discovery.
LLM-assisted writing is detectable at scale in published papers~\cite{liang2025quantifying,kobak2025delving}, increasingly peer reviews have been LLM-modified~\cite{liang2024monitoring}, and models are even proposed to generate research ideas~\cite{lu2026towards, ghareeb2026scientific}.
When LLMs help design studies, run experiments, draft manuscripts, and then review them, the models shape what the community asks, finds, and accepts~\cite{messeri2024artificial}.
The use of AI tools leads to more citations for scientists~\cite{hao2026artificial}, while this recursion risks epistemic narrowing over time~\cite{peterson2025knowledge}.
Capturing the cumulative effects of model dependence across model lifecycles will require holistic, longitudinal measurement spanning every stage of scientific discovery.

\subsection{Geopolitical Power Dynamics}

Model access is subject not only to provider decisions but also to external forces.
Export controls restrict the hardware needed to train and serve large models, and even hosted access can be withdrawn, as when the US government suspended access to Anthropic's Fable model~\cite{anthropic2026fable}.
The largest open-weight tier remains the most concentrated by institution (Figure~\ref{fig:ineq_gini_by_year}) and by country (Figure~\ref{fig:ineq_gini_by_year_country}).
Locally deployed weights cannot be deprecated, geo-blocked, or remotely altered, yet developing open models requires substantial investment in compute, data, and talent, a capacity that may become an indicator of national power.
Each country should therefore weigh building sovereign AI capability for the continuity of its science.

\begin{takeaway}{Research Recommendation}
Language models have become shared research infrastructure, yet their availability, transparency, and reach are set by a few providers and by external forces. 
The community should treat lasting, auditable, and equitable model access as a collective responsibility, secured through open alternatives and public investment, and should monitor the longitudinal impact of model dependence on scientific discovery.
\end{takeaway}

\section{Recommendations}

\subsection{Benchmarking Model Usage}
\label{checklist}

\begin{table}[t]
\centering\small
\setlength{\extrarowheight}{2.5pt}
\begin{tabularx}{\columnwidth}{@{}>{\columncolor{orange!15}[0pt][\tabcolsep]\centering\arraybackslash}p{0.55cm} X@{}}
\toprule

B1 & \textbf{Access platform and interface}: provider and endpoint/region, and the platform's API version or date header where one exists; and the interaction mode: stateless calls or chat session, noting any persistent memory, retrieval or web search, or tool use. \\
\addlinespace

B2 & \textbf{Deprecation status}: whether this model version is still served and its scheduled retirement date if announced, per the provider's model-lifecycle page, with the date the page was accessed.\\
\addlinespace

B3 & \textbf{Replicability-after-retirement plan}: released raw model outputs (e.g., cached input--output pairs); a named open-weight substitute with a reported baseline; or an explicit statement that exact replication will be impossible. \\
\addlinespace

B4 & \textbf{Usage and cost}: e.g., number of calls, total input/output tokens, and approximate spend.\\
\bottomrule
\end{tabularx}
\caption{The model-lifecycle reporting items for hosted (API) model usage.}
\label{tab:lifecycle_items}
\end{table}

\paragraph{Standardized model reporting.}

Our empirical examination reveals no standard protocol for reporting model usage, with 86\% of closed model mentions omitting an exact snapshot and 19\% of open-weight mentions stating no parameter count.
Authors routinely substitute product names for exact versions (e.g., \texttt{chatgpt-4o}), and some name their own systems after commercial products. 
The open-weight \texttt{video-chatgpt}~\cite{maaz2024video}, built on LLaVA~\cite{liu2023visual}, adopts industry naming and blurs the line between open and proprietary systems as it spreads through later publications. 
We advocate reporting each closed model at its exact snapshot and each open-weight model with its exact variant, parameter count, and any quantization or pruning applied.
Table~\ref{tab:checklist} in the Appendix gives the full checklist.

\paragraph{Model lifecycle.}

Table~\ref{tab:lifecycle_items} lists the four items we add for hosted models.
\emph{\textbf{B1}} records the access platform, without which one cannot verify which system actually produced the results.
\emph{\textbf{B2}} records the deprecation status, which all major model providers document on official lifecycle pages.
Prior work calls for documenting the exact model version and provider~\cite{feuerriegel2026reporting}.
Item \emph{\textbf{B3}} goes further and asks whether, and how, results remain reproducible once the underlying models are retired, through released outputs, a named open-weight reference, or a declared expiration.
\emph{\textbf{B4}} records usage and cost, supporting budget-aware review and the community's ability to track the price of its evidence~\cite{henderson2020towards}.

\paragraph{Keeping the checklist current.}

Reporting standards must track a moving target. 
LLM practice changes faster than the norms that regulate it, and any fixed checklist can be outdated by new technologies and experimental designs.
We believe NLP can learn from established disciplines. 
Medical reporting guidelines are versioned documents revised as methods evolve, and Institutional Review Board (IRB) style review treats oversight as an ongoing process rather than a one-time approval.
We recommend that the Responsible NLP checklist~\cite{arr2024responsible} adopt the same explicit versioning and periodic revision, especially for LLM-related items. 
Checklists can shift practice~\cite{magnusson2023checklist}, and they must not grow so demanding that they burden the resource-constrained researchers they are meant to protect.

\paragraph{Future research directions.}

Measuring published work offers important lessons about how research is produced, the focus of a field known as the science of science~\cite{fortunato2018science}.
This paper follows that tradition, but publications reveal only part of how models are accessed. 
Aggregation platforms such as OpenRouter route requests to many models through a single interface, placing research demand within a wider model access ecosystem~\cite{nagle2025latent}. 
Such intermediaries also weaken the link between the model named and the model served, as APIs that claim to offer frontier models may in fact serve a different one~\cite{zhang2026realmoney}. 
Our audit is therefore only one step toward understanding the far more complex phenomenon of LLM usage. 
Research is collaborative. 
In a multi-institution paper, the API budget or GPU cluster may come from a single research group, so including one well-resourced group can give the whole team a disproportionate advantage in model access. 
Future work can analyze acknowledgment sections as a second trace, linking model usage to the grants, compute credits, and industry programs that pay for it.

\subsection{Peer-review Practice}

The checklist serves as a codebook for review. 
Authors produce a durable record whose reproducibility status is explicit, and reviewers gain a shared basis for judging what can reasonably be asked.
Reviewers should treat the items as reporting requirements, not grounds for rejection.
A reviewer requesting additional experiments should show that the extra compute or the (often larger or paid) model is necessary to evaluate the claim rather than merely preferable, and otherwise suggest strengthening the work within its existing resources.
Cost disclosure \emph{\textbf{B4}}, in particular, must not become a discrimination channel against low-resource institutions. 
It asks for costs so that reviewers can assess dependence, not so that spending becomes an acceptance criterion.

\begin{takeaway}{Research Recommendation}
Future work should address model dependence across the whole ecosystem, from providers and access platforms to funding and peer review, and across every stage of the research process. 
Disclosure serves as a durable record of what the evidence rests on, and it should not bring backlash to low-resource communities.
\end{takeaway}

\section{Concluding Remarks}

Scientific knowledge accumulates through claims that can be re-examined, and a claim whose experiments cannot be reproduced is beyond re-examination.
With the growing use of LLMs, commercial and open-weight alike, NLP research has now become LLM-centric.
Language models touch the scientific process end to end, from idea generation and experimentation to writing and peer review, so the terms on which they are offered have become conditions of science itself.
Our audit of nine years of the ACL Anthology reveals three patterns: (i) research relies on a handful of commercial models, (ii) published evidence may expire as models are deprecated and retired, and (iii) model usage is uneven across venues, institutions, and countries.

These patterns raise a broader question of how a transformative technology diffuses.
A released model has diffused only when researchers everywhere can afford to run and build on it, and unequal access becomes unequal capability, which shapes the questions a community asks and the evidence it can produce.
As language models become instruments of scientific and economic advancement, adoption gaps that persist over the long term risk hardening into geographical disparities in global AI adoption.
If these systems are to remain this central to research, the community must address both expiring evidence and unequal access.
Keeping models accessible and claims verifiable is therefore a shared responsibility of researchers, venues, providers, and policymakers.



\section*{Limitations}

Our study has several limitations.
First, our empirical results are a lower bound on the true statistics, since some model reports are unresolvable and other commercial providers fall outside our lifecycle analysis, and the exact usage of each model (e.g., data annotation) is beyond the scope of this study.
Second, our results build on a single model's extraction, and combining extractions across models could further improve robustness; it may also miss open-weight models accessed through external APIs such as Microsoft Azure.
Third, although we advocate snapshot-level documentation of model usage, model updates remain subject to service providers, so we view our work as one step toward AI transparency rather than a complete solution.
Finally, our corpus is drawn from NLP venues and mainly covers language models; future work can extend our framework to other modalities, such as vision, speech, and robotics.

\section*{Ethical Considerations}

Our study analyzes published works through official ACL Anthology metadata and the camera-ready PDFs; no private or non-public data are used.
All findings are reported at the aggregate level of models, venues, institutions, countries, and years.
We do not name, rank, or profile individual researchers.
Author-level information is processed only to link papers to institutions and countries to avoid double counting.
All data are stored and processed locally on a secure institutional server.

Furthermore, our results document unequal access to models across venues, institutions, and countries, which reflects the field's infrastructure, not the competence of researchers or the merit of their work.
We caution against interpreting it as a quality signal, for instance when reviewers treat the absence of the newest commercial model as a weakness in itself.
In particular, resource disclosure, especially the cost item \emph{\textbf{B4}}, must not become a gatekeeping or discrimination channel in review.

\section*{Acknowledgments}

Cha was supported by the National Research Foundation of Korea grant (RS-2022-00165347).
We used Claude Code to help review code and to correct grammar.
All results and writing were produced by the authors, who take full responsibility for the content.
The authors thank Muhammad Bilal Zafar for the helpful feedback.


\bibliography{0_main}

\clearpage

\appendix

\section{ACL Anthology Dataset}
\label{sec:appendix_dataset}

\subsection{Dataset Details}
\label{sec:appendix_dataset_details}

\begin{figure}[ht]
  \centering
  \includegraphics[width=.9\linewidth]{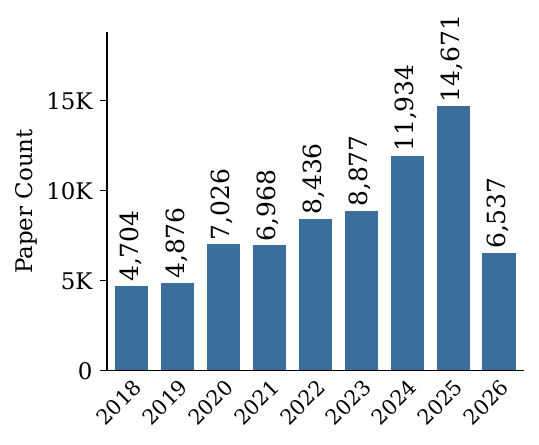}
  \caption{Yearly paper counts in the processed corpus.}
  \label{fig:papers_by_year}
\end{figure}

\noindent
All paper metadata and full texts were obtained exclusively from the ACL Anthology, the open archive for all conferences such as ACL and EMNLP, journals such as TACL, and all affiliated workshops. 
Bibliographic metadata was parsed from the Anthology's open XML metadata, taken from a fixed snapshot (2026-06-24) of the repository.\footnote{\url{https://github.com/acl-org/acl-anthology}} 
Table~\ref{tab:metadata_acl_fields} shows all available attributes from the metadata.

Restricting to papers published in 2018 or later yields 74,225 papers.
Of these, 174 papers have no fetchable PDF, and the large majority are metadata-only Anthology entries for non-archival or abstract-only content.
We extracted text from each downloaded PDF with PyMuPDF.
22 PDFs yielded no usable text, with errors such as vectorized PDFs that were run through post-processing tools and converted into images. 
After removing missing and unparseable files, \textbf{74{,}029 papers} remain with usable full text; this is the corpus used for all subsequent analyses, distributed across years as shown in Figure~\ref{fig:papers_by_year}.

\paragraph{License}

The ACL Anthology~\footnote{\url{https://aclanthology.org/faq/copyright/}} data is released under the Apache License 2.0, and all paper full texts in our corpus are licensed under Creative Commons Attribution 4.0 (CC BY).

\begin{table}[t]
\centering
\resizebox{\columnwidth}{!}{%
\begin{tabular}{ll}
\toprule
\textbf{Field} & \textbf{Notes} \\
\midrule
\texttt{anthology\_id} & e.g.\ \texttt{2023.acl-long.2} \\
\texttt{collection/volume/paper\_id}& structural keys \\
\texttt{year}/\texttt{month} & from volume \texttt{<meta>} \\
\texttt{venues} & list, e.g.\ \texttt{["acl"]} \\
\texttt{title} & \\
\texttt{authors} & list of full names \\
\texttt{abstract} &  \\
\texttt{pages} & \\
\texttt{doi} & optional \\
\texttt{bibkey} & citation key \\
\texttt{award} & \\
\texttt{booktitle} & proceedings title \\
\texttt{pdf\_url} & \url{https://aclanthology.org/<id>.pdf} \\
\bottomrule
\end{tabular}%
}
\caption{Metadata fields extracted from the Anthology XML for each paper.}
\label{tab:metadata_acl_fields}
\end{table}

\subsection{Affiliation and Country Extraction}
\label{sec:appendix_country_extraction}

The Anthology XML records author names but no affiliations.
We prompt a Gemma model with the first 2,000 characters of each paper, which cover the title, author, and affiliation header, and extract the authors with their printed affiliations and countries. 
When the header is cut off in the middle, we extend the window and rerun the extraction. 
In total, we obtain 332,008 author instances and 378,906 author–affiliation pairs, covering 73,501 of the 74,029 papers.

\paragraph{Institution and country processing.}

The 50,076 distinct raw affiliation strings are mapped to institute-level canonical names with an explicit, versioned alias list combined with the Research Organization Registry~\cite{ror}: abbreviations are expanded (e.g., mapping KAIST to Korea Advanced Institute of Science and Technology), corporate labs are rolled up to their parent brand (Microsoft Research $\rightarrow$ Microsoft, FAIR/Meta~AI $\rightarrow$ Meta, DeepMind $\rightarrow$ Google), and spelling variants are merged, yielding 18,108 distinct institutions. Institutions that share a name across countries are split by the affiliation's country (e.g., Northeastern University (US) vs.\ Northeastern University (CN)).

An affiliation's country is taken from the paper header when printed (38.5\% of affiliation instances); otherwise, it is filled in from the institution's ROR record.
After enrichment, 99.5\% of academic affiliation instances carry a country.

\subsection{Model Usage Extraction}
\label{sec:appendix_model_extraction}

\begin{table*}[t]
\centering\footnotesize
\begin{tabular}{@{}l>{\raggedright\arraybackslash}p{0.43\textwidth}>{\raggedright\arraybackslash\ttfamily}p{0.33\textwidth}@{}}
\toprule
Type & Explanation & {\normalfont Example rule} \\
\midrule
Closed (OpenAI) & API-only OpenAI models: GPT-3/4/5, ChatGPT, the o-series, and the davinci-era engines. & \textasciicircum{}gpt-?4 \\
\addlinespace
Closed (Google) & API-only Google and DeepMind models: Gemini, PaLM, Bard, Flamingo, and the closed image/video generators. & \textasciicircum{}gemini \\
\addlinespace
Closed (Anthropic) & Anthropic's Claude family; the boundary excludes lookalike names such as CLAUDETTE. & (\textasciicircum{}|\allowbreak [ /])claude(\$|\allowbreak [ .\textbackslash{}-\textbackslash{}d]) \\
\addlinespace
Closed (other) & Closed models of all other commercial vendors: Grok-3 and later, Qwen-Max, GLM-4 API tiers, Doubao, Nova, \dots & \textasciicircum{}qwen[\textbackslash{}d\textbackslash{}.]*[ -]?(vl-)?(max|\allowbreak plus|\allowbreak turbo) \\
\addlinespace
Open LLM & Open-weight LLMs and VLMs: Llama, Mistral, Qwen, DeepSeek, Gemma, GPT-2-successors such as OPT and BLOOM, instruction-tuned T5s, LLaVA, \dots & \textasciicircum{}(meta[ -]?)?(open-?)?llama \\
\addlinespace
Classic PLM & BERT-era pretrained models: BERT and its derivatives (RoBERTa, SciBERT, \dots). & bert \\
\addlinespace
Other model & Pretrained non-text-LM systems: speech, vision, diffusion, and OCR models, plus commercial translation/ASR services. & \textasciicircum{}stable[ -]?diffusion \\
\addlinespace
Method/metric & Not a model: metrics (BLEU, BERTScore), training methods (LoRA, DPO), architectures (LSTM), and toolkits (spaCy). & \textasciicircum{}lora|\allowbreak \textasciicircum{}qlora|\allowbreak \textasciicircum{}dpo\$|\allowbreak \textasciicircum{}ppo\$|\allowbreak \textasciicircum{}rlhf \\
\addlinespace
Ambiguous & Bare product-level names that cannot be resolved between open- and closed-access versions. & \textasciicircum{}gpt\$ \\
\addlinespace
Unclassified & Fallback for names that match no rule (a long tail of one-off systems); excluded from the analyses as noise. & {\normalfont ---\ (no rule)} \\
\bottomrule
\end{tabular}
\caption{The ten model-type categories.}
\label{tab:group_overview}
\end{table*}

\paragraph{Extraction criterion and output.}

We extract model names using Gemma to return every model or system a paper itself \emph{runs}, covering proposed models, fine-tuned backbones, prompted or queried APIs, and evaluated baselines, across commercial LLM APIs, open-weight LLMs, and pre-LLM-era pretrained models. 
Models named only as cited or prior work are excluded after checking the surrounding context of every occurrence, as are training and optimization methods (PPO, DPO, LoRA), datasets, benchmarks, and metrics. 
Each extracted item is a JSON object \texttt{\{name, version, size, evidence\}} with the name exactly as written, the version and parameter size included only when the paper states them and never inferred, and a contiguous evidence quote that must contain the name.

We then normalize the verbatim names. 
Pure spelling variants are unified via Unicode normalization, casefolding, and unification of dash and space separators, so that ``GPT 4o'', gpt-4o, and gpt4o become one name displayed as its most frequent corpus spelling, while sizes, tiers, versions, and modalities always stay distinct (gpt-4o-mini $\neq$ gpt-4o). 
Within each paper, a generic name is absorbed by a more specific name of the same model when its tokens form a strict prefix of the specific name and every extra token is a pure specifier, either a date pin (\texttt{0806}, \texttt{2024-08-06}) or a parameter size (\texttt{7b}, \texttt{8x7b}). 
Any other extra token (\texttt{mini}, \texttt{flash}, \texttt{instruct}, \texttt{vl}) changes the model and blocks absorption. 
Each paper thus contributes a list of distinct models.


An ordered rule set assigns every distinct model name to one of ten groups covering closed models of OpenAI, Google, and Anthropic, closed models of other commercial providers, open-weight LLMs, BERT-era pretrained models, and noise (non-language services, method or metric names, and unresolvable names). 
Rules for closed API-only tiers precede their open-weight siblings (Gemini before Gemma, Qwen-Max before Qwen), and grouping follows access rather than vendor, so GPT-2 counts as open-weight. 
Timestamped variants additionally carry a base name with date pins removed (\texttt{gpt-4o-2024-08-06} $\rightarrow$ \texttt{gpt-4o}) for reporting, while lifecycle matching (Appendix~\ref{sec:appendix_lifecycle}) runs on the full pinned name. 
Table~\ref{tab:group_overview} gives one representative rule per group.

\paragraph{Computation}

All experiments were run locally on a GPU cluster instance equipped with four NVIDIA H200 NVL GPUs (141\,GB HBM3e) and loaded via the Hugging Face Transformers library.
All extraction was performed with \texttt{google/gemma-4-31B-it} (bf16) served by vLLM, and the memory fits the entire paper body in a single 65,536-token context window.

\paragraph{Software}

The pipeline is implemented in Python~3.13. 
Anthology metadata is parsed with \texttt{lxml}~6.1; 
PDFs are fetched with \texttt{requests}~2.34 and their text extracted with PyMuPDF~1.27. 
LLM extraction is served with vLLM~0.23.0 on Transformers~5.12.1 and PyTorch~2.11. All processing, statistics, and figures use pandas~3.0, NumPy~2.4, SciPy~1.15 (OLS regressions via \texttt{scipy.stats}), and Matplotlib~3.11.


\subsection{Commercial Model Lifecycle}
\label{sec:appendix_lifecycle}


We compiled Table~\ref{tab:model_lifecycle} from official provider documentation only (accessed 2026-07-22). 
OpenAI and Anthropic each maintain a single deprecations page that lists, for every model version, the deprecation announcement and the retirement date; all their dates are taken verbatim from those pages. 
For Google, the deprecations page records shutdown dates for Gemini~2.0 onward, while announcement dates appear in the official release notes; the PaLM~2 and Gemini~1.x generations have been removed from the live pages, so their dates (marked $^{\approx}$ in the table) follow the Vertex~AI model-lifecycle documentation as published at the time. 
The Released column uses the snapshot date embedded in the API identifier where one exists.
For all analyses, a version's status (alive, deprecated, or retired) is evaluated as of the corpus snapshot date (2026-06-24).

\section{Usage Measurements}
\label{sec:appendix_gini}

\paragraph{Per-paper usage flags.} 

For each paper $p$ and model type $t\in\{\textsc{closed},\textsc{open},\textsc{plm}\}$, we set the binary flag $u_p(t)\in\{0,1\}$: whether $p$ used at least one closed LLM from the three major model providers, at least one open-weight LLM, or at least one classic PLM, respectively. 
The \textsc{open} flag is refined into size tiers ($\leq$10B, 10--70B, $>$70B) based on the extracted parameter sizes.

\paragraph{Why binary flags}

All usage rates build on the at-least-one flags $u_p(t)\in\{0,1\}$, so a paper counts once for a model type regardless of how many models of that type it used. 
We deliberately do not weight by model counts for three reasons. 
First, mention-weighted rates would largely track paper genre, since a benchmarking paper contains many systems and size variants, while an application paper may use a single model. 
Second, count-based measures also inherit every normalization and within-paper deduplication choice, whereas the at-least-one flag is invariant to them.
Third, our question concerns the reach of each model type, and binary flags give a clear signal of the share of papers that used that model type. 
How many models a paper uses is a complementary quantity that we track separately.

\paragraph{Usage rates by venue.} 
Venues partition the corpus since every paper belongs to exactly one venue, so venue-level rates use whole-paper counting: for venue $v$ with paper set $\mathcal{P}_v$, the adoption rate of type $t$ is the share of its papers using that type. \[ r_v(t)=\frac{1}{|\mathcal{P}_v|}\sum_{p\in\mathcal{P}_v}u_p(t) \]

\paragraph{Usage rates by country.} 

Institutions and countries do not partition papers, and paper credit is split fractionally because 58\% of papers list authors from more than one institution.
A paper $p$ with author set $A_p$ gives each author $1/|A_p|$ of one unit of credit, and an author $a$ with $k_a$ resolvable affiliations passes $1/(|A_p|\,k_a)$ to each affiliated institution and that institution's country. 
With $w_p(c)\in[0,1]$ the total credit of paper $p$ landing on country $c$, the country's usage rate for type $t$ is the credit-weighted share, which measures a country's participation in papers that use type $t$:
\[ r_c(t)=\frac{\sum_p w_p(c)\,u_p(t)}{\sum_p w_p(c)} \]

\paragraph{Usage concentration: Gini index.} 

The same credit split applies at the institutional level, where the weight $w_p(i)\in[0,1]$ is paper $p$'s fractional author-share credit at institution $i$. 
Institution $i$'s usage mass for type $t$ in year $y$ is the total credit it receives from that year's papers that use that type, \[ x_i(t,y)=\sum_{p\,:\,\mathrm{year}(p)=y} w_p(i)\,u_p(t). \] With $\mathcal{I}_y$ the set of institutions receiving any credit in year $y$, $n=|\mathcal{I}_y|$, and $\bar{x}$ the mean mass, the per-year Gini index of type $t$ is \[ G_y(t)=\frac{\sum_{i}\sum_{j}|x_i-x_j|}{2n^{2}\bar{x}}, \] 
$G=0$ when every publishing institution carries an equal share of type-$t$ usage, and $G\rightarrow1$ as usage concentrates in a single institution. 
The concentration patterns remain consistent when we add controls such as venue and team size.

\section{National Wealth and Model Usage}
\label{sec:appendix_gdp}

\paragraph{GDP data.}
We measure national wealth as GDP per capita at purchasing power parity in constant 2021 international dollars (World Bank World Development Indicators, bulk release of 2026-07-13, accessed 2026-08-02). This is the standard measure for cross-country wealth comparisons.

\paragraph{Estimation.} 

For each publication year and model type, we fit the bivariate OLS regression \[ \log_{10}\mathrm{GDPpc}_c=\alpha+\beta\cdot \frac{u_c(t,y)}{10\,\mathrm{pp}}+\varepsilon_c , \] regressing wealth on the usage share in 10-percentage-point units, where $u_c(t,y)$ is the share of country $c$'s papers in year $y$ that use model type $t$. 
To keep the usage shares stable, a country enters the usage-share regressions for year $y$ only with $\geq$10 fractional papers in that year.
Classic-PLM regressions cover every year from 2018 to 2025. 
Closed and open-weight LLM usage is negligible before 2023, so those regressions are restricted to 2023--2025. 
GDP per capita is log-transformed because it is approximately log-normal across countries, and the 10-pp unit only rescales the coefficient without affecting fit statistics or $p$-values. 
Figure~\ref{fig:gdp_coefficient} and Table~\ref{tab:gdp_regressions} report the percent difference in GDP per capita associated with a 10-pp-higher usage share. 
Table~\ref{tab:gdp_papercount} details the percent difference in GDP per capita associated with paper counts.
The patterns still hold when we add controls such as team size and adjust for multiple comparisons.
Note that all estimates are correlational, not causal effects.

\begin{figure}[t]
  \centering
  \includegraphics[width=.88\linewidth]{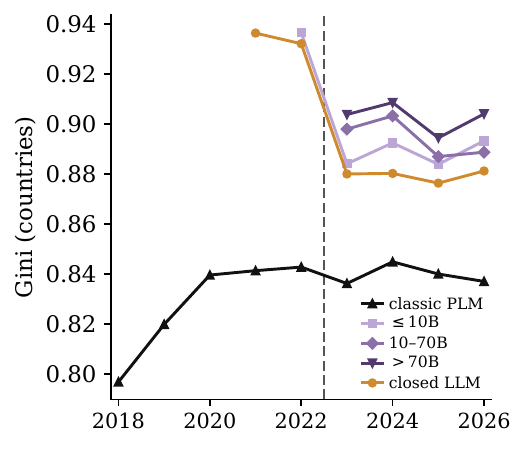}
  \caption{Concentration of model-type usage across publishing countries, measured by the per-year Gini index. Open-weight LLMs are split by parameter size. The dashed line separates the pre- and post-LLM eras.}
  \label{fig:ineq_gini_by_year_country}
\end{figure}

\begin{table}[ht]
\centering
\small
\begin{tabular}{lrrrr}
\toprule
\textbf{Year} & $N$ & \textbf{Closed} & \textbf{Open} & \textbf{PLM} \\
\midrule
\multicolumn{5}{l}{\emph{TACL (journal)}} \\
2023 & 98 & 18 & 13 & 89 \\
2024 & 95 & 49 & 48 & 78 \\
2025 & 82 & 60 & 74 & 55 \\
2026 & 34 & 71 & 88 & 59 \\
\textbf{2023--26} & 309 & 45 & 49 & 73 \\
\midrule
\multicolumn{5}{l}{\emph{CL (journal)}} \\
2023 & 30 & 13 & 7 & 67 \\
2024 & 41 & 44 & 27 & 71 \\
2025 & 38 & 55 & 53 & 68 \\
2026 & 11 & 36 & 82 & 82 \\
\textbf{2023--26} & 120 & 39 & 35 & 70 \\
\midrule
\multicolumn{5}{l}{\emph{ACL / EMNLP / NAACL (main conferences)}} \\
2023 & 2,430 & 33 & 23 & 82 \\
2024 & 3,139 & 62 & 67 & 59 \\
2025 & 4,955 & 65 & 82 & 44 \\
2026 & 2,650 & 65 & 88 & 31 \\
\textbf{2023--26} & 13,174 & 58 & 69 & 52 \\
\bottomrule
\end{tabular}
\caption{Model-type adoption rates (\%) by publication year for the two journals (TACL, CL) and the three main conferences.}
\label{tab:app_venue_delay}
\end{table}

\section{Checklist}
\label{app:appendix_checklist}

This section presents the complete proposed model-usage reporting checklist (Table~\ref{tab:checklist}). 
The checklist is completed once per model usage. 
Part~A applies to every usage, and the access gate then assigns hosted usages to Part~B and locally run open-weight usages to Part~C. 
Table~\ref{tab:checklist_mapping} compares the checklist against the existing major reporting standards, namely GUIDE-LLM~\cite{feuerriegel2026reporting}, the ACL Responsible NLP Checklist~\cite{arr2024responsible}, the NeurIPS paper checklist~\cite{neurips2026checklist}, and the AAAI reproducibility checklist~\cite{aaai2026reproducibility}. 
Most Part~A items are at least partially covered by one or more of these standards, whereas the lifecycle items B2 (deprecation status) and B3 (replicability-after-retirement plan) have no counterpart.

We further provide two worked examples. 
Table~\ref{tab:checklist_example} completes the checklist for this work's own model usage, a single open-weight model run from local weights, so the access gate selects Part~C and the pinned checkpoint revision, environment, and compute budget make our extraction runs exactly re-executable. 
Table~\ref{tab:checklist_example_closed} shows a synthetic example, a hypothetical study running a hosted closed model, where the gate selects Part~B and the lifecycle items record a model already scheduled for shutdown (B2) along with its replicability-after-retirement plan (B3).

\begin{table*}[t]
\centering\small
\setlength{\extrarowheight}{2.5pt}
\newcommand{\ck}{\ensuremath{\Box}}
\newcommand{\loc}{\makebox[\linewidth]{\dotfill}}
\newcommand{\banner}[2]{\multicolumn{3}{@{}l@{}}{\hspace*{0.3cm}\colorbox{#1}{\parbox{\dimexpr\textwidth-0.6cm-2\fboxsep\relax}{#2}}}}
\newcommand{\idcell}[2]{\multicolumn{1}{@{\hspace{\dimexpr0.3cm+\tabcolsep\relax}}>{\columncolor{#1}[\tabcolsep][\tabcolsep]\centering\arraybackslash}p{0.55cm}}{#2}}
\begin{tabularx}{\textwidth}{@{\hspace{\dimexpr0.3cm+\tabcolsep\relax}}p{0.55cm} X p{2.3cm}@{\hspace{\dimexpr0.3cm+\tabcolsep\relax}}}
\toprule
& \textbf{Item} & \textbf{Where (\S\,/\,N/A)}\\
\midrule

\banner{black!10}{\textbf{Part A. Every model usage} \emph{(always complete)}}\\
\idcell{black!10}{A1} & \textbf{Model name and version}: exact model name \emph{and} version/snapshot (e.g.\ \texttt{gpt-4o-2024-08-06}), not a family name (``GPT-4'') & \loc\\
\idcell{black!10}{A2} & \textbf{Role(s) of the model in the study}: e.g., proposed system, fine-tuned backbone, evaluated baseline, evaluator or judge, data generation or annotation & \loc\\
\idcell{black!10}{A3} & \textbf{Access date}: the date the model was run or accessed (identically named endpoints drift over time) & \loc\\
\idcell{black!10}{A4} & \textbf{Prompts}: full system prompt, templates, few-shot exemplars --- or, for programmatic or agentic pipelines, the code and configuration that generate them --- released with the paper, or a stated reason they cannot be released & \loc\\
\idcell{black!10}{A5} & \textbf{Decoding and request parameters}: e.g., temperature, top-$p$, maximum tokens, number of samples per input, and random seed where supported & \loc\\
\idcell{black!10}{A6} & \textbf{Runs and variance}: number of runs per configuration, and the aggregation and variance reported (mean/std, or a statement that a single run was used) & \loc\\
\midrule

\multicolumn{3}{@{}p{\dimexpr\textwidth-4\tabcolsep\relax}@{}}{\textbf{Access gate.} This usage was \quad \ck~\colorbox{orange!15}{hosted (API or chat interface) $\rightarrow$ Part~B} \quad \ck~\colorbox{violet!14}{run from local weights $\rightarrow$ Part~C}}\\
\midrule

\banner{orange!15}{\textbf{Part B. Hosted-access usage} \emph{(provider-served, including open-weight models behind a third-party API)}}\\
\idcell{orange!15}{B1} & \textbf{Access platform and interface}: provider and endpoint/region, and the platform's API version or date header where one exists; and the interaction mode: stateless calls or chat session, noting any persistent memory, retrieval or web search, or tool use & \loc\\
\idcell{orange!15}{B2} & \textbf{Deprecation status}: whether this model version is still served and its scheduled retirement date if announced, per the provider's model-lifecycle page, with the date the page was accessed. & \loc\\
\idcell{orange!15}{B3} & \textbf{Replicability-after-retirement plan}: released raw model outputs (e.g., cached input--output pairs); a named open-weight substitute with a reported baseline; or an explicit statement that exact replication will be impossible. 
& \loc\\
\idcell{orange!15}{B4} & \textbf{Usage and cost}: e.g., number of calls, total input/output tokens, and approximate spend & \loc\\
\addlinespace

\banner{violet!14}{\textbf{Part C. Locally run open-weight usage}}\\
\idcell{violet!14}{C1} & \textbf{Compute and hardware}: GPU/accelerator type and count, and compute budget (e.g., GPU-hours) & \loc\\
\idcell{violet!14}{C2} & \textbf{Pinned environment}: exact checkpoint revision, numerical precision/quantization, and inference framework \emph{and version}. If fine-tuned: availability of the training data and the resulting checkpoint & \loc\\
\bottomrule
\end{tabularx}
\caption{The proposed model-usage reporting checklist, completed once per \emph{model usage}: every usage fills Part~A, and the access gate then assigns Part~B (hosted) or Part~C (locally run).}
\label{tab:checklist}
\end{table*}

\begin{table*}[t]
\centering\small
\setlength{\extrarowheight}{1.5pt}
\setlength{\tabcolsep}{14pt}
\newcommand{\pp}{\textsuperscript{\,p}}
\newcommand{\idb}[2]{{\setlength{\fboxsep}{1.5pt}\colorbox{#1}{\makebox[0.5cm]{\strut #2}}}}
\begin{tabular}{@{}l l l l l@{}}
\toprule
\textbf{Ours} & \textbf{GUIDE-LLM} & \textbf{ACL RNC} & \textbf{NeurIPS} & \textbf{AAAI}\\
\midrule
\idb{black!10}{A1} name + version & B.1 & B1\pp & Q12\pp & ---\\
\idb{black!10}{A2} role & A.1\pp & --- & Q16\pp & ---\\
\idb{black!10}{A3} access date & B.1 & --- & --- & ---\\
\idb{black!10}{A4} prompts & C.1--C.2 & --- & Q5\pp & E3--E4\pp\\
\idb{black!10}{A5} decoding params & B.3 & C2\pp & Q6\pp & E6, E12\pp\\
\idb{black!10}{A6} runs, variance & B.3\pp & C3 & Q7 & E9--E10\\
\addlinespace
\idb{orange!15}{B1} platform, mode & B.2, B.4, B.5 & C4\pp & Q12\pp & ---\\
\idb{orange!15}{\textbf{B2}} \textbf{deprecation status} & \textbf{---} & \textbf{---} & \textbf{---} & \textbf{---}\\
\idb{orange!15}{\textbf{B3}} \textbf{post-retirement plan} & \textbf{---} & \textbf{---} & \textbf{---} & \textbf{---}\\
\idb{orange!15}{B4} usage, cost & opt. & C1\pp & Q8\pp & ---\\
\addlinespace
\idb{violet!14}{C1} compute, hardware & opt. & C1 & Q8 & E7\pp\\
\idb{violet!14}{C2} pinned environment & B.1, B.3\pp & C4\pp & Q5, Q12\pp & E7\pp\\
\bottomrule
\end{tabular}
\caption{Mapping of our items onto GUIDE-LLM~\cite{feuerriegel2026reporting}, the ACL Responsible Checklist~\cite{arr2024responsible}, the NeurIPS checklist~\cite{neurips2026checklist} (Q$n$: question $n$), and the AAAI checklist~\cite{aaai2026reproducibility} (E$n$: computational item $n$). ``---'': no counterpart; $^{\textrm{p}}$: partial coverage; ``opt.'': optional GUIDE-LLM item.}
\label{tab:checklist_mapping}
\end{table*}

\begin{table*}[t]
\centering\small
\setlength{\extrarowheight}{2.5pt}
\newcommand{\exbanner}[2]{\multicolumn{2}{@{}l@{}}{\hspace*{0.3cm}\colorbox{#1}{\parbox{\dimexpr\textwidth-0.6cm-2\fboxsep\relax}{#2}}}}
\newcommand{\exidcell}[2]{\multicolumn{1}{@{\hspace{\dimexpr0.3cm+\tabcolsep\relax}}>{\columncolor{#1}[\tabcolsep][\tabcolsep]\centering\arraybackslash}p{0.55cm}}{#2}}
\begin{tabularx}{\textwidth}{@{\hspace{\dimexpr0.3cm+\tabcolsep\relax}}p{0.55cm} X@{\hspace{\dimexpr0.3cm+\tabcolsep\relax}}}
\toprule
& \textbf{This work's answers}\\
\midrule

\exbanner{black!10}{\textbf{Part A. Every model usage} --- example for one model used in this paper, \texttt{google/gemma-4-31B-it}}\\
\exidcell{black!10}{A1} & \textbf{Model name and version}: \texttt{google/gemma-4-31B-it} (its only public release); the exact checkpoint revision is pinned under C2\\
\exidcell{black!10}{A2} & \textbf{Role(s)}: extraction instrument (data annotation): identifies the models each paper uses and the paper's author affiliations from its text\\
\exidcell{black!10}{A3} & \textbf{Access date}: all extraction runs executed in July 2026, on the 2026-06-24 corpus snapshot\\
\exidcell{black!10}{A4} & \textbf{Prompts}: both system prompts released verbatim in Appendix.\\
\exidcell{black!10}{A5} & \textbf{Decoding and request parameters}: greedy decoding (temperature 0), at most 4{,}096 output tokens, 65{,}536-token context window; serving-framework defaults otherwise\\
\exidcell{black!10}{A6} & \textbf{Runs and variance}: one greedy pass per paper; empty or unparsable outputs re-run once with identical settings\\
\midrule

\multicolumn{2}{@{}p{\dimexpr\textwidth-4\tabcolsep\relax}@{}}{\textbf{Access gate.} \quad $\boxtimes$~\colorbox{violet!14}{run from local weights $\rightarrow$ Part~C} \quad (no hosted usage; Part~B does not apply)}\\
\midrule

\exbanner{violet!14}{\textbf{Part C. Locally run open-weight usage}}\\
\exidcell{violet!14}{C1} & \textbf{Compute and hardware}: single-GPU vLLM jobs on NVIDIA H200 NVL (141\,GB) nodes of a SLURM cluster; $\approx$90 GPU-hours in total, including repeated and failed jobs\\
\exidcell{violet!14}{C2} & \textbf{Pinned environment}: checkpoint revision \texttt{518276fb130d}, bf16 weights; vLLM 0.23.0, Transformers 5.12.1; the model was not fine-tuned\\
\bottomrule
\end{tabularx}
\caption{Worked example: the checklist of Table~\ref{tab:checklist} completed for one example model used in this work. The model shown is an open-weight model run from local weights, so the access gate selects Part~C and Part~B is empty.}
\label{tab:checklist_example}
\end{table*}

\begin{table*}[t]
\centering\small
\setlength{\extrarowheight}{2.5pt}
\newcommand{\synbanner}[2]{\multicolumn{2}{@{}l@{}}{\hspace*{0.3cm}\colorbox{#1}{\parbox{\dimexpr\textwidth-0.6cm-2\fboxsep\relax}{#2}}}}
\newcommand{\synidcell}[2]{\multicolumn{1}{@{\hspace{\dimexpr0.3cm+\tabcolsep\relax}}>{\columncolor{#1}[\tabcolsep][\tabcolsep]\centering\arraybackslash}p{0.55cm}}{#2}}
\begin{tabularx}{\textwidth}{@{\hspace{\dimexpr0.3cm+\tabcolsep\relax}}p{0.55cm} X@{\hspace{\dimexpr0.3cm+\tabcolsep\relax}}}
\toprule
& \textbf{A hypothetical closed model study's answers}\\
\midrule

\synbanner{black!10}{\textbf{Part A. Every model usage} --- the study used exactly one model, \texttt{gpt-4o-2024-05-13}}\\
\synidcell{black!10}{A1} & \textbf{Model name and version}: \texttt{gpt-4o-2024-05-13}\\
\synidcell{black!10}{A2} & \textbf{Role(s)}: zero-shot and few-shot classification of stance in online political discussions\\
\synidcell{black!10}{A3} & \textbf{Access date}: April 2026\\
\synidcell{black!10}{A4} & \textbf{Prompts}: zero-shot and few-shot prompts in Appendices~B and~C, respectively; few-shot exemplars available through the accompanying code and sample data in the anonymous repository; the complete dataset will be released after peer review\\
\synidcell{black!10}{A5} & \textbf{Decoding and request parameters}: temperature 1, top-$p$ 0.95, at most 50 output tokens\\
\synidcell{black!10}{A6} & \textbf{Runs and variance}: five runs per configuration; \S4.1 reports the mean accuracy and F1 with 95\% confidence intervals across runs\\
\midrule

\multicolumn{2}{@{}p{\dimexpr\textwidth-4\tabcolsep\relax}@{}}{\textbf{Access gate.} \quad $\boxtimes$~\colorbox{orange!15}{hosted (API or chat interface) $\rightarrow$ Part~B} \quad (no local-weights usage; Part~C does not apply)}\\
\midrule

\synbanner{orange!15}{\textbf{Part B. Hosted-access usage}}\\
\synidcell{orange!15}{B1} & \textbf{Access platform and interface}: OpenAI Chat Completions API (version 2020-10-01; United States region); stateless requests without persistent memory, retrieval, web search, or tool use\\
\synidcell{orange!15}{B2} & \textbf{Deprecation status}: still served, but scheduled to shut down on October~23, 2026, per \url{https://developers.openai.com/api/docs/deprecations} (accessed 2026-08-02)\\
\synidcell{orange!15}{B3} & \textbf{Replicability-after-retirement plan}: all raw model outputs (cached input--output pairs) are released with the code and data; after the scheduled shutdown, exact re-execution will no longer be possible through any official interface\\
\synidcell{orange!15}{B4} & \textbf{Usage and cost}: 45{,}000 calls; $\sim$60 input and $\sim$10 output tokens per call; approximate cost US\$11.25\\
\bottomrule
\end{tabularx}
\caption{Synthetic worked example: the checklist of Table~\ref{tab:checklist} completed for a hypothetical study that uses a single hosted closed model, so the access gate selects Part~B and Part~C is empty. The lifecycle facts in B2 are real as of the stated access date; all other entries are illustrative.}
\label{tab:checklist_example_closed}
\end{table*}

\begin{figure}[ht]
  \centering
  \includegraphics[width=\linewidth]{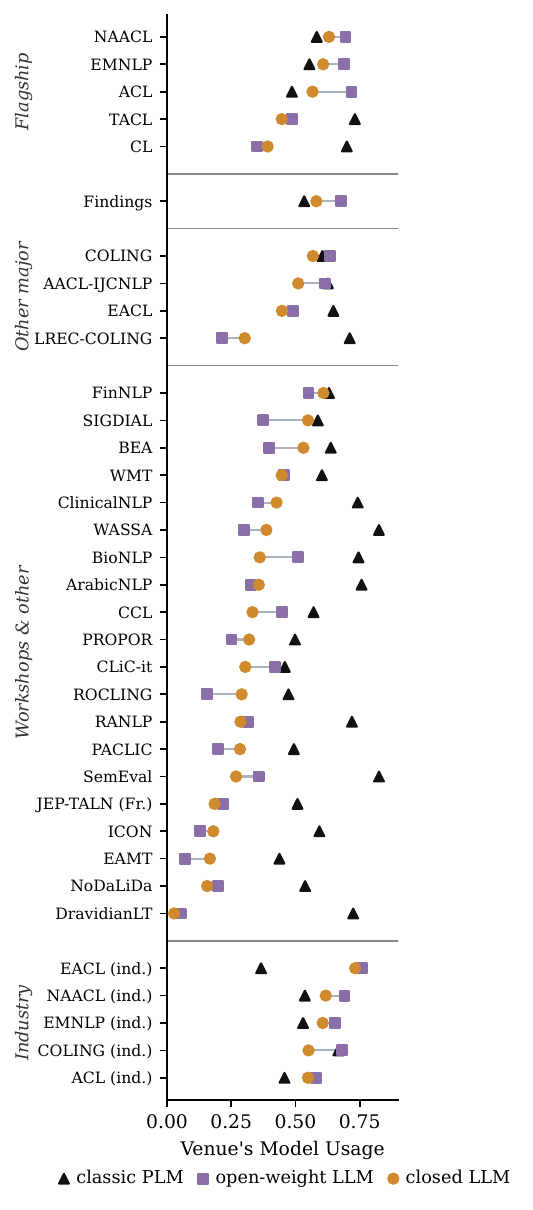}
  \caption{Share of each venue's papers model usage.}
  \label{fig:app_ineq_venue}
\end{figure}

\begin{table}[ht]
\centering
\resizebox{\columnwidth}{!}{%
\begin{tabular}{llrrcr}
\toprule
\textbf{Type} & \textbf{Year} & $n$ & \textbf{\%$\Delta$} & \textbf{95\% CI} & $p$ \\
\midrule
PLM & 2018 & 36 & $-5.2$ & $[-25.7,\,+21.1]$ & 0.663 \\
PLM & 2019 & 35 & $+0.3$ & $[-19.2,\,+24.3]$ & 0.981 \\
PLM & 2020 & 42 & $-14.6$ & $[-24.4,\,-3.6]$ & 0.012 \\
PLM & 2021 & 41 & $-7.6$ & $[-23.8,\,+12.0]$ & 0.410 \\
PLM & 2022 & 49 & $-4.8$ & $[-16.3,\,+8.3]$ & 0.444 \\
PLM & 2023 & 45 & $+2.1$ & $[-11.2,\,+17.3]$ & 0.768 \\
PLM & 2024 & 53 & $-8.8$ & $[-22.7,\,+7.6]$ & 0.271 \\
PLM & 2025 & 59 & $-26.1$ & $[-35.7,\,-15.0]$ & $<$0.001 \\
\midrule
Open & 2023 & 45 & $+33.2$ & $[+1.3,\,+75.0]$ & 0.041 \\
Open & 2024 & 53 & $+20.4$ & $[+6.9,\,+35.7]$ & 0.003 \\
Open & 2025 & 59 & $+31.6$ & $[+17.1,\,+47.9]$ & $<$0.001 \\
\midrule
Closed & 2023 & 45 & $+21.0$ & $[-0.2,\,+46.7]$ & 0.052 \\
Closed & 2024 & 53 & $+26.5$ & $[+11.7,\,+43.2]$ & $<$0.001 \\
Closed & 2025 & 59 & $+14.2$ & $[-1.8,\,+32.9]$ & 0.083 \\
\bottomrule
\end{tabular}%
}
\caption{Percent difference in GDP per capita per 10-pp increase in model usage share, by year and model type, with 95\% CIs and two-sided $p$-values. Classic PLM is estimated for all years with published GDP (2018--2025); open-weight and closed LLMs only for 2023--2025. The 2023--2025 rows underlie Figure~\ref{fig:gdp_coefficient}.}
\label{tab:gdp_regressions}
\end{table}

\begin{table}[ht]
\centering
\resizebox{\columnwidth}{!}{%
\begin{tabular}{lrrcr}
\toprule
\textbf{Year} & $n$ & \textbf{\%$\Delta$} & \textbf{95\% CI} & $p$ \\
\midrule
2018 & 81 & $+49.5$ & $[+20.0,\,+86.3]$ & $<$0.001 \\
2019 & 87 & $+56.1$ & $[+29.6,\,+88.0]$ & $<$0.001 \\
2020 & 93 & $+58.7$ & $[+32.0,\,+90.7]$ & $<$0.001 \\
2021 & 89 & $+50.7$ & $[+25.8,\,+80.4]$ & $<$0.001 \\
2022 & 99 & $+63.4$ & $[+38.9,\,+92.2]$ & $<$0.001 \\
2023 & 95 & $+65.1$ & $[+38.4,\,+97.1]$ & $<$0.001 \\
2024 & 112 & $+88.2$ & $[+61.9,\,+118.7]$ & $<$0.001 \\
2025 & 109 & $+76.4$ & $[+50.3,\,+107.0]$ & $<$0.001 \\
\bottomrule
\end{tabular}%
}
\caption{Percent difference in GDP per capita per tenfold increase in total fractional paper counts, with 95\% CIs and two-sided $p$-values. Counts include all papers regardless of model usage.}
\label{tab:gdp_papercount}
\end{table}

\begin{table*}[t]
\centering
\footnotesize
\begin{tabular}{lrrrr}
\toprule
\textbf{Country/Region} & $N$ & \textbf{Closed} & \textbf{Open} & \textbf{PLM} \\
\midrule
China & 8,306 & 55.5 & 68.3 & 51.1 \\
United States & 7,374 & 56.4 & 56.5 & 55.6 \\
Germany & 2,405 & 33.6 & 43.2 & 62.6 \\
United Kingdom & 1,655 & 43.7 & 51.2 & 65.3 \\
India & 1,596 & 32.2 & 38.7 & 68.0 \\
France & 1,332 & 24.9 & 33.8 & 61.1 \\
South Korea & 1,224 & 62.8 & 65.2 & 54.8 \\
Japan & 1,206 & 41.2 & 39.7 & 56.2 \\
Canada & 866 & 48.3 & 48.3 & 59.5 \\
Italy & 861 & 32.3 & 43.9 & 51.7 \\
Singapore & 660 & 64.2 & 69.6 & 53.6 \\
Hong Kong & 621 & 61.1 & 74.3 & 44.8 \\
Netherlands & 599 & 31.5 & 40.0 & 62.1 \\
Switzerland & 592 & 42.5 & 40.0 & 57.9 \\
Australia & 559 & 55.1 & 55.6 & 58.0 \\
Spain & 458 & 27.7 & 39.6 & 65.6 \\
Brazil & 381 & 29.0 & 23.4 & 50.5 \\
United Arab Emirates & 348 & 60.7 & 63.9 & 55.3 \\
Israel & 326 & 47.4 & 50.8 & 65.7 \\
Taiwan & 323 & 44.5 & 40.6 & 55.2 \\
Bangladesh & 263 & 23.7 & 31.4 & 80.3 \\
Denmark & 256 & 27.5 & 37.6 & 67.1 \\
Sweden & 237 & 23.8 & 27.6 & 59.8 \\
Ireland & 219 & 34.3 & 33.8 & 56.8 \\
Czechia & 216 & 24.7 & 35.9 & 56.4 \\
Vietnam & 216 & 42.6 & 45.9 & 69.3 \\
Belgium & 209 & 30.2 & 29.0 & 60.0 \\
Finland & 190 & 30.4 & 23.2 & 46.4 \\
Russia & 189 & 36.7 & 58.9 & 68.5 \\
Portugal & 180 & 31.4 & 38.5 & 55.0 \\
Poland & 163 & 33.6 & 38.1 & 59.4 \\
Iran & 138 & 43.4 & 44.7 & 71.9 \\
Norway & 136 & 20.4 & 31.6 & 56.0 \\
Austria & 126 & 37.8 & 43.0 & 61.5 \\
Romania & 123 & 35.3 & 42.2 & 68.3 \\
Saudi Arabia & 123 & 51.2 & 44.6 & 57.8 \\
Qatar & 118 & 54.1 & 51.8 & 66.0 \\
\bottomrule
\end{tabular}
\hfill
\begin{tabular}{lrrrr}
\toprule
\textbf{Country/Region} & $N$ & \textbf{Closed} & \textbf{Open} & \textbf{PLM} \\
\midrule
Turkey & 109 & 38.5 & 39.6 & 68.7 \\
Greece & 91 & 46.5 & 45.2 & 68.5 \\
Egypt & 83 & 27.8 & 39.3 & 72.2 \\
Mexico & 78 & 17.8 & 17.0 & 65.3 \\
Pakistan & 62 & 46.7 & 33.7 & 72.3 \\
Bulgaria & 59 & 34.6 & 24.5 & 42.9 \\
Thailand & 56 & 49.2 & 45.1 & 58.3 \\
Macao & 56 & 52.1 & 67.3 & 36.6 \\
Luxembourg & 56 & 40.4 & 41.4 & 62.2 \\
Estonia & 53 & 38.7 & 26.0 & 63.4 \\
Philippines & 50 & 21.6 & 19.8 & 44.4 \\
South Africa & 48 & 14.1 & 12.6 & 48.3 \\
Iceland & 47 & 33.6 & 23.8 & 57.6 \\
Slovenia & 47 & 18.7 & 22.6 & 64.1 \\
Slovakia & 44 & 49.6 & 60.8 & 81.8 \\
New Zealand & 43 & 41.5 & 37.7 & 64.6 \\
Sri Lanka & 41 & 16.1 & 24.2 & 69.0 \\
Ukraine & 40 & 46.2 & 48.1 & 68.5 \\
Croatia & 35 & 29.4 & 34.9 & 71.4 \\
Indonesia & 34 & 39.0 & 47.8 & 68.4 \\
Hungary & 32 & 19.4 & 23.2 & 63.0 \\
Morocco & 31 & 35.2 & 34.4 & 64.8 \\
Nepal & 30 & 24.4 & 14.8 & 57.7 \\
Nigeria & 28 & 26.7 & 28.3 & 67.5 \\
Palestine & 26 & 34.8 & 18.1 & 71.7 \\
Colombia & 24 & 57.4 & 31.8 & 70.2 \\
Latvia & 23 & 29.4 & 26.4 & 43.3 \\
Argentina & 22 & 34.9 & 44.5 & 53.9 \\
Chile & 21 & 35.4 & 30.1 & 66.7 \\
Malta & 18 & 16.2 & 14.0 & 75.3 \\
Malaysia & 17 & 50.2 & 47.5 & 66.9 \\
Algeria & 17 & 11.5 & 14.4 & 86.7 \\
Serbia & 17 & 32.6 & 19.6 & 37.4 \\
Tunisia & 14 & 12.3 & 8.5 & 79.8 \\
Ethiopia & 14 & 16.2 & 11.0 & 83.2 \\
Kazakhstan & 12 & 33.7 & 27.7 & 70.0 \\
Uruguay & 11 & 27.8 & 47.9 & 82.9 \\
\bottomrule
\end{tabular}
\caption{Country/region-level model-type adoption rates (2023--2026; fractional author-share counting). \textbf{Closed}, \textbf{Open}, and \textbf{PLM} give the share (\%) of a country's papers using at least one closed LLM from the three major providers, open-weight LLM, or classic PLM, respectively. Shown are all 74 countries/regions with fractional paper count $N \geq 10$, sorted by $N$.}
\label{tab:country_rates_full}
\end{table*}

\begin{figure*}[t]
\centering
\begin{tcolorbox}[colback=gray!5!white, colframe=gray!100!black, width=\linewidth]
\begin{lstlisting}[style=prompt, basicstyle=\footnotesize]
You are an information-extraction system. You will be given the raw text of the beginning of a scientific paper (extracted automatically from an ACL Anthology PDF). Your task is to extract the paper's authors, in order, and the affiliation(s) of each author, including the affiliation's country when it is printed.

PROPERTIES OF THE INPUT
- The text is raw PDF-extracted text: line breaks can occur anywhere (including inside names), and the reading order of header elements may be slightly scrambled.
- The header typically contains, in some order: a proceedings/copyright banner (e.g. "Proceedings of the ... pages ..."), the paper title, author names, affiliations, email addresses, and the beginning of the Abstract.
- Author-affiliation links are often marked with superscript markers (digits 1,2,3 or symbols *, (dagger), (double-dagger)) placed after author names and before affiliation lines. Resolve these markers to assign the right affiliation(s) to the right author.
- Sometimes affiliations appear directly under a group of author names without any markers; in that case every author in that group shares those affiliations.
- The PDF text may contain typographic artifacts: detached accents ("Saarbrucken" for "Saarbruecken", "Benoit" for "Benoit") and ligature glyphs ("Artificial" for "Artificial"). Repair these to the proper characters in your output.

OUTPUT RULES
1. Output ONLY a single JSON object. No markdown code fences, no explanations, no extra text.
2. Use exactly this schema:
   {"authors": [{"name": "<full name>", "affiliations": [{"institution": "<full affiliation>", "country": "<country or null>"}, ...]}, ...]}
3. Preserve the author order as printed in the paper.
4. "name" is the full author name verbatim (after artifact repair), WITHOUT superscript markers, footnote symbols, email addresses, or parenthetical additions such as pronouns.
5. Each "institution" is ONE complete affiliation. When one affiliation spans several consecutive lines (department/lab, institution, city, country), JOIN them into a single string separated by ", " -- do not emit the lines as separate affiliations. Example: "Department of Computational Linguistics" / "Saarland University" / "Saarbruecken, Germany" becomes "Department of Computational Linguistics, Saarland University, Saarbruecken, Germany". IMPORTANT: never join text belonging to DIFFERENT superscript markers -- each marker starts a new, separate affiliation. Only join continuation lines that carry no marker of their own (they extend the affiliation opened by the most recent marker), or a shared umbrella line as in rule 6. An author whose name carries several markers gets several separate affiliation objects, one per marker.
6. If a university or organization line is printed as a shared umbrella for several department/institute lines (e.g. "1Institute for X; 2Department of Y" followed by "Stony Brook University"), append the umbrella to EACH of those affiliations: "Institute for X, Stony Brook University" and "Department of Y, Stony Brook University".
7. "country" is the country of that affiliation exactly as printed in the header (e.g. "Germany", "USA", "P.R. China", "United Kingdom"). If no country is printed for that affiliation, use null. NEVER infer the country from world knowledge -- only copy what is printed.
8. Never include email addresses in any field.
9. If an author's affiliation cannot be determined from the text, use an empty list []. Never guess or invent an affiliation.
10. Include only the authors of THIS paper -- not editors, program committee members, cited authors, or names from the proceedings banner.
11. If the excerpt contains no identifiable author block, output {"authors": []}.

EXAMPLE:
Input--Output pairs
\end{lstlisting}
\end{tcolorbox}
\caption{System prompt for extracting authors, their affiliations, and printed country from each paper header.}
\label{fig:prompt_author}
\end{figure*}

\begin{figure*}[t]
\centering
\begin{tcolorbox}[colback=gray!5!white, colframe=gray!100!black, width=\linewidth]
\begin{lstlisting}[style=prompt, basicstyle=\scriptsize]
You are an expert NLP research assistant building a dataset of which models each paper actually RUNS in its own experiments. Capture the named foundation/commercial LLMs a paper runs or compares against as carefully as its own proposed systems. When several models appear in one list or table, record each one separately -- never collapse them. Copy names and evidence verbatim, and make every evidence quote literally contain the model's name; never invent anything that is not in the provided text.

The user message contains ONLY an excerpt of one research paper. From that excerpt, list every model or system that THIS paper RUNS in its own experiments -- anything the authors run, prompt, query, fine-tune, train, evaluate, or execute as a baseline. Whether it is the paper's own proposed model, a backbone it builds on, or a comparison baseline makes NO difference -- include all of them, on equal footing. If the excerpt shows the authors ran a model but you are not fully certain, still include it; only leave out models that clearly appear as cited prior work / general discussion only.

INCLUDE (do not skip any):
- EVERY named foundation/commercial LLM it runs or compares against -- e.g. GPT-3.5, GPT-4, GPT-4o, o1, Gemini, Claude, Llama, Mistral, Qwen, DeepSeek, T5, BERT. Capture these even when they appear ONLY as rows in a results/baseline table among many domain-specific system names -- never drop them in favour of the novel system names. Capture the most SPECIFIC version string the text gives (e.g. gpt-4-0613, gpt-3.5-turbo-0301, text-davinci-003, Claude-3.5-Sonnet) -- exact API snapshot names matter downstream.
- every named OPEN-WEIGHT model it runs, however niche -- e.g. Llama-3-8B, Qwen2.5, Yi, OLMo, MiniCPM, ModernBERT, BERT-base. Do not drop a model for being small, old, or unfamiliar.
- every named PRE-TRAINED language model of the pre-LLM era it runs or fine-tunes -- e.g. BERT, RoBERTa, DeBERTa, ELECTRA, XLNet, ALBERT, BART, mBART, T5, mT5, XLM-R, GPT-2, DistilBERT, SpanBERT, ERNIE. These count exactly like LLMs.
- any other named model or system the paper proposes, fine-tunes, or compares against

ENUMERATE EXHAUSTIVELY -- do NOT collapse lists or tables:
- When models appear in a comma- or "and"-separated list, or as the rows/columns of a results/baseline table, output EVERY item as its own object. Never summarize, never emit just one representative, never fold variants into a family.
- Treat each named version as a DISTINCT item: e.g. GPT-4o, GPT-4o-mini, GPT-4.1, GPT-4.1-mini and GPT-5 are FIVE separate entries; Llama-3-8B and Llama-3-70B are two.
- Worked example: the sentence "we evaluate GPT-4o-mini, GPT-4o, GPT-4.1-mini, GPT-4.1, and GPT-5" must yield FIVE objects, one per model -- not a single GPT-4/GPT-5 entry.

CHECK THE SURROUNDING CONTEXT OF EVERY MENTION:
A model is usually mentioned several times in one paper (an introduction citation, the experimental setup, a results table, an appendix). Before deciding on any model, look at the text SURROUNDING EACH of its mentions:
- Usage: include the model if ANY mention's surrounding context shows the authors running it, even when its other mentions are pure citations; leave it out only if EVERY mention is citation/discussion-only.
- Evidence: take the quote from a mention whose surrounding context shows the paper running/evaluating the model (setup, results, comparison) -- never from an introduction/related-work citation of it.
- Version and size: harvest these from the context around ANY of the mentions -- papers often state them exactly once (an experimental-setup sentence, a table row, an appendix list) even when the name appears many times elsewhere.
- Attribution: when several models share one sentence or table row, make sure a nearby version/size/date actually belongs to THIS model and not to its neighbour.

EXCLUDE (do NOT output these):
- training/optimization/fine-tuning methods or algorithms -- e.g. PPO, GRPO, DPO, RLHF, SFT, LoRA, DoRA, ReFT, RLAIF, ORPO, KTO. These are methods, NOT models; never list them.
- models named ONLY as cited/related/prior work or in passing discussion, with no sign in this excerpt that the authors ran them. Introductions and related-work sections cite many models -- a citation like "LLMs such as GPT-4 (OpenAI, 2023) have advanced NLP" is NOT usage. But if the same passage shows the authors running it ("we prompt GPT-4"), include it.
- datasets, benchmarks, metrics, evaluation toolkits

For each item, return a JSON object with exactly these keys:
  "name": the name EXACTLY as written in the text -- do NOT rename, expand, normalize, or add/remove hyphens, spaces, or case
  "version": the model's version/variant identifier IF the excerpt states one -- an API snapshot or date string (e.g. "0613", "gpt-4-0613", "2024-08-06", "20240229"), a release number (e.g. "2", "3.5", "v1.5", "2.5"), or a named variant (e.g. "turbo", "mini", "Sonnet", "Instruct") -- copied verbatim from the excerpt. The version often appears NEAR the name rather than inside it (e.g. "we use the June 2023 snapshot of GPT-4") -- capture it from there too. Use null if the excerpt states no version. NEVER infer a version from outside knowledge. A CITATION YEAR is NOT a version: in "DeepFix (Gupta et al., 2017)" the 2017 is the cited paper's publication year -- version stays null.
  "size": the parameter-count/scale variant IF the excerpt states one -- e.g. "7B", "70B", "175B", "8x7B", "base", "large", "xxl" -- copied verbatim (from the name, the surrounding text, or a table). NOT an embedding/hidden dimension ("768-dimensional" is a dimension, not a size -- use null). Use null if not stated. NEVER infer it from outside knowledge.
  "evidence": a SHORT (<=200 char) quote that is ONE SINGLE CONTIGUOUS span COPY-PASTED verbatim from the excerpt -- NEVER two fragments joined with "..." or "...", never paraphrased, nothing skipped in the middle. It MUST literally contain the "name" string above and show the paper running/evaluating it. When the name sits in a long list or table row, quote the SHORTEST contiguous span that contains the name -- the quote may start and end mid-sentence. Do NOT use a figure or table caption, or any span, that does not itself contain the name.

Before you finish, re-scan the excerpt for any mention of foundation/commercial LLM families and make sure every specific version the paper runs is in your list. This includes, but is NOT limited to: GPT / ChatGPT / o1 / o3 / o4 (OpenAI), Gemini / PaLM / Gemma (Google), Claude (Anthropic), Phi (Microsoft), Grok (xAI), Llama (Meta), Mistral / Mixtral, Qwen (Alibaba), DeepSeek, Command / Command-R (Cohere), Yi (01.AI), GLM / ChatGLM (Zhipu), Falcon (TII), Nemotron (NVIDIA), Granite (IBM), OLMo (AI2), InternLM, Baichuan, BERT / T5. Capture models from ANY company -- include named LLMs even if their vendor is not in this list.

Return ONLY a JSON array of such objects (use [] if none). No prose, no markdown.
\end{lstlisting}
\end{tcolorbox}
\caption{System prompt for extracting the models each paper runs.}
\label{fig:prompt_model}
\end{figure*}

\begin{table*}[t]
\centering
\footnotesize
\begin{tabular}{@{}llll@{}}
\toprule
\textbf{Model (API identifier)} & \textbf{Released} & \textbf{Deprecation announced} & \textbf{Retired} \\
\midrule
\multicolumn{4}{@{}l}{\textit{\textbf{OpenAI}}} \\
\texttt{text-davinci-002}            & 2022-03$^{\approx}$ & 2023-07-06 & 2024-01-04 \\
\texttt{code-davinci-002}            & 2022-03$^{\approx}$ & 2023-03-20 & 2023-03-23 \\
\texttt{text-davinci-003}            & 2022-11-28 & 2023-07-06 & 2024-01-04 \\
\texttt{gpt-3.5-turbo-0301}          & 2023-03-01 & 2023-06-13 & 2024-09-13 \\
\texttt{gpt-4-0314}                  & 2023-03-14 & 2023-06-13 & 2026-03-26 \\
\texttt{gpt-4-32k}                   & 2023-03-14 & 2024-06-06 & 2025-06-06 \\
\texttt{gpt-3.5-turbo-0613}          & 2023-06-13 & 2023-11-06 & 2024-09-13 \\
\texttt{gpt-4-0613}                  & 2023-06-13 & 2026-04-22 & 2026-10-23$^{\dagger}$ \\
\texttt{gpt-3.5-turbo-1106}          & 2023-11-06 & 2025-09-26 & 2026-09-28$^{\dagger}$ \\
\texttt{gpt-4-1106-preview}          & 2023-11-06 & 2025-09-26 & 2026-10-23$^{\dagger}$ \\
\texttt{gpt-4-vision-preview}        & 2023-11-06 & 2024-06-06 & 2024-12-06 \\
\texttt{gpt-3.5-turbo-0125}          & 2024-01-25 & 2026-04-22 & 2026-10-23$^{\dagger}$ \\
\texttt{gpt-4-0125-preview}          & 2024-01-25 & 2025-09-26 & 2026-03-26 \\
\texttt{gpt-4-turbo-2024-04-09}      & 2024-04-09 & 2026-04-22 & 2026-10-23$^{\dagger}$ \\
\texttt{gpt-4o-2024-05-13}           & 2024-05-13 & 2026-04-22 & 2026-10-23$^{\dagger}$ \\
\texttt{gpt-4o-mini-2024-07-18}      & 2024-07-18 & ---        & --- \\
\texttt{gpt-4o-2024-08-06}           & 2024-08-06 & ---        & --- \\
\texttt{o1-preview}                  & 2024-09-12 & 2025-04-28 & 2025-07-28 \\
\texttt{o1-mini}                     & 2024-09-12 & 2025-04-28 & 2025-10-27 \\
\texttt{gpt-4o-2024-11-20}           & 2024-11-20 & ---        & --- \\
\texttt{o1-2024-12-17}               & 2024-12-17 & 2026-04-22 & 2026-10-23$^{\dagger}$ \\
\texttt{o3-mini-2025-01-31}          & 2025-01-31 & 2026-04-22 & 2026-10-23$^{\dagger}$ \\
\texttt{gpt-4.5-preview}             & 2025-02-27 & 2025-04-14 & 2025-07-14 \\
\texttt{o3-2025-04-16}               & 2025-04-16 & 2026-06-11 & 2026-12-11$^{\dagger}$ \\
\texttt{gpt-5-2025-08-07}            & 2025-08-07 & 2026-06-11 & 2026-12-11$^{\dagger}$ \\
\midrule
\multicolumn{4}{@{}l}{\textit{\textbf{Anthropic}}} \\
\texttt{claude-1.x} / \texttt{claude-instant-1.x} & 2023-03-14 & 2024-09-04 & 2024-11-06 \\
\texttt{claude-2.0}                  & 2023-07-11 & 2025-01-21 & 2025-07-21 \\
\texttt{claude-2.1}                  & 2023-11-21 & 2025-01-21 & 2025-07-21 \\
\texttt{claude-3-sonnet-20240229}    & 2024-02-29 & 2025-01-21 & 2025-07-21 \\
\texttt{claude-3-opus-20240229}      & 2024-02-29 & 2025-06-30 & 2026-01-05 \\
\texttt{claude-3-haiku-20240307}     & 2024-03-07 & 2026-02-19 & 2026-04-20 \\
\texttt{claude-3-5-sonnet-20240620}  & 2024-06-20 & 2025-08-13 & 2025-10-28 \\
\texttt{claude-3-5-sonnet-20241022}  & 2024-10-22 & 2025-08-13 & 2025-10-28 \\
\texttt{claude-3-5-haiku-20241022}   & 2024-10-22 & 2025-12-19 & 2026-02-19 \\
\texttt{claude-3-7-sonnet-20250219}  & 2025-02-19 & 2025-10-28 & 2026-02-19 \\
\texttt{claude-sonnet-4-20250514}    & 2025-05-14 & 2026-04-14 & 2026-06-15 \\
\texttt{claude-opus-4-20250514}      & 2025-05-14 & 2026-04-14 & 2026-06-15 \\
\texttt{claude-opus-4-1-20250805}    & 2025-08-05 & 2026-06-05 & 2026-08-05$^{\dagger}$ \\
\midrule
\multicolumn{4}{@{}l}{\textit{\textbf{Google}}} \\
PaLM~2 (\texttt{text-/chat-bison})   & 2023-05-10 & 2024-04-09$^{\approx}$ & 2025-04-21$^{\approx}$ \\
\texttt{gemini-1.0-pro}              & 2023-12-13 & 2024-09-24$^{\approx}$ & 2025-04-21$^{\approx}$ \\
\texttt{gemini-1.5-pro-001}          & 2024-05-23 & 2024-09-24$^{\approx}$ & 2025-05-24$^{\approx}$ \\
\texttt{gemini-1.5-pro/flash-002}    & 2024-09-24 & 2025-04-29$^{\approx}$ & 2025-09-24$^{\approx}$ \\
\texttt{gemini-2.0-flash-001}        & 2025-02-05 & 2026-02-18 & 2026-06-01 \\
\texttt{gemini-2.5-pro}              & 2025-06-17 & 2025-06-17             & 2026-10-16$^{\dagger}$ \\
\texttt{gemini-2.5-flash}            & 2025-06-17 & 2025-06-17             & 2026-10-16$^{\dagger}$ \\
\texttt{gemini-2.5-flash-lite}       & 2025-07-22 & 2025-07-22             & 2026-10-16$^{\dagger}$ \\
\bottomrule
\end{tabular}

\vspace{4pt}
\begin{minipage}{0.98\textwidth}
\footnotesize
\emph{Sources} (accessed 2026-07-22):\newline
OpenAI: \url{https://developers.openai.com/api/docs/deprecations}\newline
Anthropic: \url{https://platform.claude.com/docs/en/about-claude/model-deprecations}\newline
Google:~\url{https://ai.google.dev/gemini-api/docs/deprecations}\newline
\hphantom{Google:~} \url{https://ai.google.dev/gemini-api/docs/changelog}
\end{minipage}
\caption{Lifecycle of the commercial (closed, API-only) model versions of OpenAI, Anthropic, and Google used in ACL-Anthology papers: release date, deprecation announcement, and retirement (shutdown) date. ``---'' = no deprecation announced as of 2026-07-22. $^{\dagger}$~scheduled, still serving on 2026-07-22; $^{\approx}$~not on the live pages cited below, from archived Vertex~AI documentation or conventional dating.}
\label{tab:model_lifecycle}
\end{table*}


\end{document}